%% file: main.tex
\documentclass[preprint,12pt]{elsarticle}

\usepackage{xcolor}
\usepackage{subfigure}
\usepackage{amssymb}
\usepackage{amsmath}
\usepackage{url}
\usepackage{comment}

\usepackage[noend]{algpseudocode}
\usepackage{algorithm}
\usepackage{tabularx}

\journal{Nuclear Physics B}

\begin{document}

\begin{frontmatter}



\title{A Self-Adjusting FEM–BEM Coupling Scheme for the Nonlinear Poisson–Boltzmann Equation}

\author[USM]{Mauricio Guerrero-Montero}
\author[Kin]{Micha\l~ Bosy}
\author[USM,CCTVAL]{Christopher D. Cooper}
\affiliation[USM]{organization={Department of Mechanical Engineering, Universidad Técnica Federico Santa María},
             addressline={Av. España 1680},
             city={Valparaíso},
             postcode={2390123},
             state={Valparaíso},
             country={Chile}}

\affiliation[Kin]{organization={School of Computer Science and Mathematics, Kingston University London},
             addressline={55-59 Penrhyn Road},
             city={Kingston Upon Thames},
             postcode={KT1 2EE},
             state={London},
             country={UK}}
\affiliation[CCTVAL]{organization={Centro Científico Tecnológico de Valparaíso (CCTVal), Universidad Técnica Federico Santa María},
             addressline={Av. España 1680},
             city={Valparaíso},
             postcode={2390123},
             state={Valparaíso},
             country={Chile}}



\begin{abstract}
The Poisson-Boltzmann equation is widely used to model molecular electrostatics; however, it is usually solved in linearised form because the sinh nonlinearity is challenging, limiting its applicability in highly charged systems such as nucleic acids. This work presents a solution method for the nonlinear Poisson-Boltzmann equation based on a coupled finite/boundary element scheme that automatically finds an optimal relaxation parameter, ensuring fast and reliable convergence of the nonlinear solver without user intervention. We validated our solver against APBS for a spherical cavity, and used RNA-based structures to perform a thorough study of the different algorithmic choices, and to test our implementation. We found that the best alternative to solve the Poisson-Boltzmann equation was using a Newton-Raphson method where the nonlinearity was gradually introduced with a cubic approximation in the first iteration. Newton-Raphson was also the best method to find the optimal relaxation factor, reducing the number of iterations by $\sim$40\%. Including other optimisation techniques, we were able to obtain a 1.37$\times$ speed-up with respect to the best hand-picked relaxation factor for 1HC8 (molecule with highest charge in our tests), avoiding any trial-and-error process to find the relaxation factor.
\end{abstract}

\begin{graphicalabstract}
\includegraphics[width=0.8\textwidth]{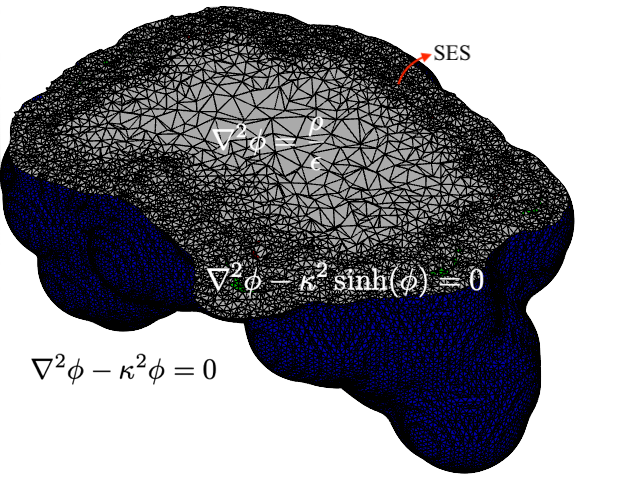}
\end{graphicalabstract}



\end{frontmatter}

\include{Introduccion_Principal}

\include{Metodologia_Principal}

\include{Resultados_Principal}

\include{Conclusiones_Principal}

\appendix

\include{Appendices}

\section*{Acknowledgements}
M.G-M. acknowledges the support from Universidad Técnica Federico Santa María through funding from the Programa de Iniciación a la Investigación Científica (PIIC). C.D.C. acknowledges the support from CCTVal through ANID PIA/
APOYO AFB220004








\bibliographystyle{apsrev}
\bibliography{main}   

\end{document}

%% file: Introduccion_Principal.tex
\section{Introduction}

The accurate computation of electrostatic interactions plays a critical role in the analysis of biomolecular systems. Among the available theoretical frameworks, the Poisson–Boltzmann equation (PBE) stands out as a widely used continuum model for describing the electrostatics of solvated biomolecules \cite{shen1995calculation,Baker2004,fu2022accurate}.  This equation accounts for the spatial distribution of mobile ions in an electrolyte and the dielectric contrast between the solute and the solvent. The PBE is widely used in linear form due to simplicity and lower computational cost; however, it is limited to scenarios involving low electrostatic potentials. In contrast, in nonlinear form the PBE captures the ionic response near highly charged interfaces more accurately, but solving it efficiently remains a computational challenge. 

The PBE has been solved numerically using finite differences (FDM) \cite{GilsonETal1985,BakerETal2001,rocchia2001extending,boschitsch2011fast,JurrusETal2018}, finite elements (FEM) \cite{cortis1997numerical,chen2007finite,xie2007new,bond2010first}, boundary elements (BEM) \cite{shaw1985theory,yoon1990boundary,juffer1991electric,boschitsch2002fast,LuETal2006,geng2013treecode,cooper2014biomolecular,search2022towards}, (semi) analytical \cite{felberg2017pb,siryk2022arbitrary,jha2023linear,jha2023domain}, and hybrid approaches \cite{boschitsch2004hybrid,ying2018hybrid,bsbbbc2023coupling}.
Conventional methods typically adopt a two-region model, in which the solute and solvent domains are separated by a dielectric interface. While this model allows for efficient and accurate boundary integral formulations, BEM is limited to the linear PBE \cite{boschitsch2002fast,altman2009accurate}. To solve this, we develop an extended three-region model to enable a hybrid formulation in which the nonlinear PBE is solved only in a near-solute region with a finite element method, where the higher electrostatic potential makes nonlinearities more important, while retaining a linear description with boundary integrals in the outer domain for computational efficiency.

The hyperbolic sine nonlinearity is a difficulty because it generates a highly stiff system that affects convergence.\cite{nicholls1991rapid} This has been solved using relaxation and Newton-type methods \cite{boschitsch2004hybrid,li2020newton,holst1995numerical,cai2010performance}, with contradicting claims in which SOR is sometimes more efficient \cite{li2020newton} whereas Newton iterations are better in other cases \cite{holst1995numerical,cai2010performance}. In addition,  convergence is extremely sensitive to the relaxation factor ($\omega$). Consequently, users often resort to trial-and-error to select an appropriate $\omega$, a process that is both inefficient and unreliable. Existing automatic schemes tend to involve procedures with multiple algorithmic steps, which complicate their practical implementation~\cite{holst1995numerical}.

In this work, we propose a coupled finite element–boundary element (FEM–BEM) formulation featuring a nonlinear solver that automatically adjusts the relaxation factor during iterations, ensuring fast and reliable convergence. The FEM–BEM coupling provides a natural division of labour: the finite element region captures the nonlinear response, while the boundary element formulation efficiently represents the solvent further away. The relaxation factor adjustment is easy to implement and is not limited to the FEM-BEM framework.

Validation against APBS benchmarks and tests on highly charged systems demonstrate that the method delivers both accuracy and computational efficiency, confirming its potential as a versatile and reliable tool for nonlinear electrostatic modeling.
Beyond the PBE, the proposed framework extends naturally to other nonlinear systems governed by Laplace operators with reaction terms, e.g. reaction–diffusion and nonlinear Helmholtz equations. Even without employing coupling, our method can be readily adapted to other problems because all nonlinearities are handled entirely within the FEM domain.

%% file: Metodologia_Principal.tex
\newcommand{\Ma}{\mathbf{A}} 
\newcommand{\Vb}{\mathbf{b}} 
\newcommand{\Vcc}{\mathbf{N}} 

\newcommand{\Vc}{\Vcc\left(\Phi_{L}+\Phi_{NL}^{k}\right)}  
\newcommand{\Vdcf}{\partial_{\Phi}\Vcc\left(\Phi_{L}+\Phi_{NL}^{k}\right)} 
\newcommand{\Vdc}{\partial_{\omega_{k}}\Vcc\left(\Phi_{L}+\Phi_{NL}^{k}\right)} 
\newcommand{\Vddc}{\partial_{\omega_{k}}^{2}\Vcc\left(\Phi_{L}+\Phi_{NL}^{k}\right)} 
\newcommand{\Vdddc}{\partial_{\omega_{k}}^{3}\Vcc\left(\Phi_{L}+\Phi_{NL}^{k}\right) } 

\newcommand{\za}{\mathbf{p}} 
\newcommand{\zb}{\mathbf{r}(\omega_{k})} 
\newcommand{\zdb}{\partial_{\omega_{k}}\mathbf{r}(\omega_{k})} 
\newcommand{\zc}{\Delta_k(\Phi_{NL})}
\newcommand{\zddb}{\partial_{\omega_{k}}^{2}\mathbf{r}(\omega_{k})} 
\newcommand{\zd}{\mathbf{d}(\Phi_{NL}^{k}(\omega_{k}))} 
\newcommand{\zdd}{\partial_{\omega_{k}}\zd} 

\section{Methodology}
\subsection{Model Problem}
\label{sec:sharp_inter}
The Poisson–Boltzmann equation (PBE) is given by:
\begin{equation} 
    \label{eq:EcGeneral_PBE}
    -\nabla\cdot(\epsilon(\mathbf{x})\nabla\phi(\mathbf{x}))+\bar{\kappa}^{2}(\mathbf{x})N(\phi(\mathbf{x}))= \rho(\mathbf{x}),
\end{equation}
where $\epsilon(\mathbf{x})$ is the relative permittivity, $\rho(\mathbf{x})$ is the charge density, $\bar{\kappa}(\mathbf{x})$ is the inverse Debye–Hückel length, and $\phi(\mathbf{x})$ is the electrostatic potential. The term $N(\phi(\mathbf{x}))$ specifies what kind of equation we consider (linear or nonlinear PBE). 

\begin{figure}[h]
\centering
\includegraphics[width=0.5\textwidth]{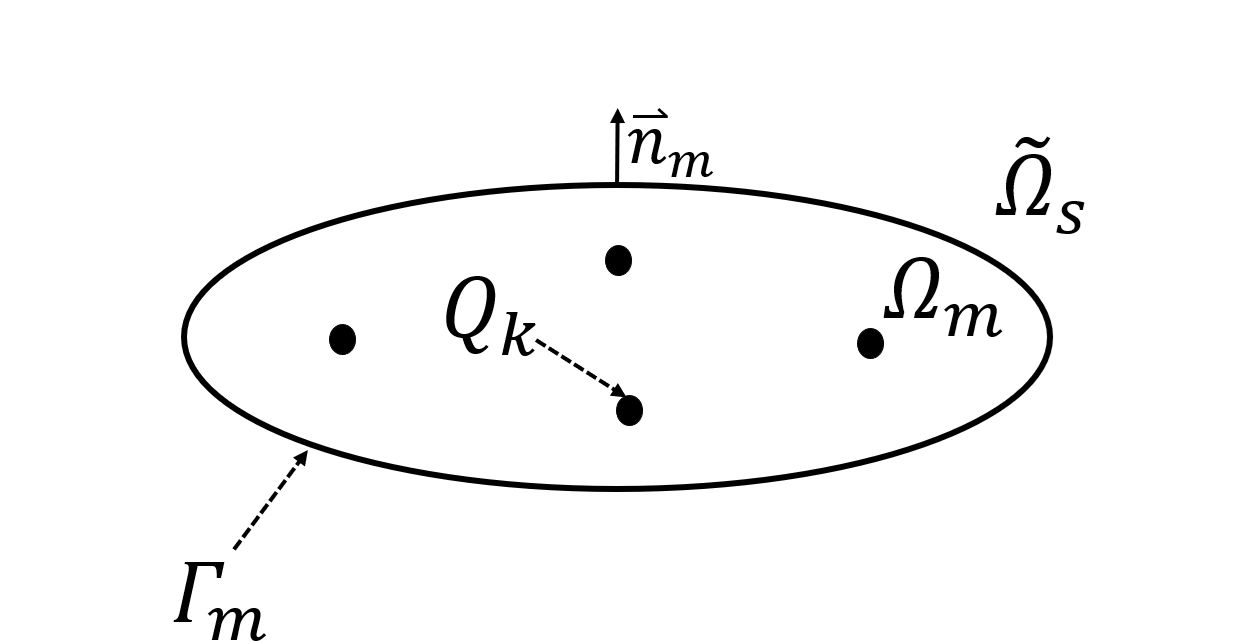}
\caption{Two-region domain split according to Equation \eqref{eq:Variables}.}
\label{fig:1_Iterfaz}
\end{figure}

Figure \ref{fig:1_Iterfaz} illustrates the most common application of the PBE to the molecular solvation problem, where the system is divided into two regions: the solute domain $\Omega_m$ and the solvent domain $\widetilde{\Omega_{s}}$, separated by a sharp dielectric interface $\Gamma_m$. The physical parameters are then defined as: 
\begin{align}
\label{eq:Variables}
    \epsilon(\mathbf{x}) = 
\left\{\begin{matrix} 
     \epsilon_{m} & \mathbf{x}\in\Omega_{m} \\ 
     \epsilon_{s} & \mathbf{x}\in \widetilde{\Omega_{s}}
\end{matrix}\right., &\quad
 \rho(\mathbf{x}) = 
\left\{\begin{matrix} 
     \rho_c & \mathbf{x}\in\Omega_{m} \\ 
     0  &  \mathbf{x}\in \widetilde{\Omega}_c
\end{matrix}\right. , \\ \nonumber
\bar{\kappa}^{2}(\mathbf{x}) = 
\left\{\begin{matrix} 
     0 & \mathbf{x} \ \in \ \Omega_{m}  \\  
     \epsilon_{s}\kappa_{s}^{2} & \mathbf{x}\in \widetilde{\Omega_{s}}
\end{matrix}\right.,&\quad
N(\phi(\mathbf{x})) = \begin{cases}
    sinh(\phi(\mathbf{x})) & \text{ for nonlinear PBE} \\
    \phi(\mathbf{x}) & \text{ for linear PBE}  
    \end{cases}
\end{align}
 where $\rho_{c} = \sum_{k=1}^{N_{q}}Q_{k}\delta\left (\mathbf{x}-\mathbf{x}_{k} \right )$ represents a distribution of $n_c$ delta-like charges located at $\mathbf{x}_k$.

\begin{figure}[h]
\centering
\includegraphics[width=0.5\textwidth]{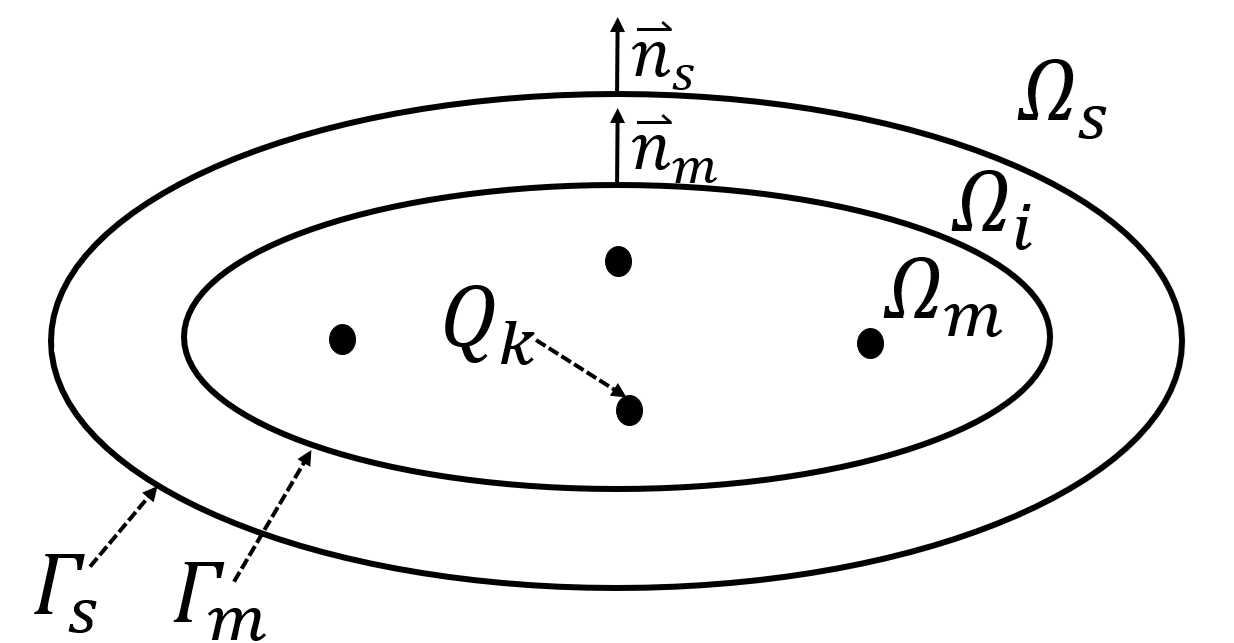}
\caption{Domain split into three regions according to Equation \eqref{eq:Variables_2Superficie}.}
\label{fig:2_Iterfaces}
\end{figure}

The nonlinear effects of the PBE are localised in high-potential regions of $\widetilde{\Omega}_s$, which should occur near the molecular surface. To account for this behaviour accurately, Figure~\ref{fig:2_Iterfaces} introduces an extended three-region formulation for PBE where $\widetilde{\Omega}_s$ is split into $\Omega_{i}$, an interface region between the solvent and solute, similar to the Stern layer region discussed by Altman {\it et al.} \cite{altman2009accurate}, and $\Omega_{s}$ in the far field. 
These regions are separated by two interfaces, denoted $\Gamma_{m}$, between $\Omega_m$ and $\Omega_i$,  and $\Gamma_{s}$, between $\Omega_i$ and $\Omega_s$. The physical parameters in Equation~\eqref{eq:Variables} are still valid, and we can distinguish 
\begin{align}
\label{eq:Variables_2Superficie}
\phi(\mathbf{x}) = 
\left\{\begin{matrix} 
     \phi_{m}(\mathbf{x})  &  \mathbf{x}\in\Omega_{m} \\ 
     \phi_{i}(\mathbf{x})  &  \mathbf{x}\in\Omega_{i} \\ 
     \phi_{s}(\mathbf{x})  &  \mathbf{x}\in\Omega_{s}  
\end{matrix}\right. ,
&& \widetilde{N}(\phi(\mathbf{x})) = 
\left\{\begin{matrix} 
     0 & \mathbf{x}\in\Omega_{m} \\ 
     N(\phi(\mathbf{x}))  &  \mathbf{x}\in\Omega_{i} \\ 
     \phi(\mathbf{x})  &  \mathbf{x}\in\Omega_{s}  
\end{matrix}\right. .
\end{align} 

Substituting parameters from Equation~\eqref{eq:Variables_2Superficie} into Equation~\eqref{eq:EcGeneral_PBE} yields the following system of partial differential equations, along with the appropriate interface conditions: 
\begin{equation} 
\label{eq:pbe_vp}
\left\{\begin{matrix} 
     -\nabla^{2} \epsilon_{m}(\mathbf{x}) \phi_{m}(\mathbf{x})= \rho_{c} & \mathbf{x} \ \in \ \Omega_{m} \\[2mm]
     -\epsilon_{s} \nabla^{2} \phi_{i}(\mathbf{x})+\epsilon_{s}\kappa_{s}^{2}N(\phi_{i}(\mathbf{x}))
     =0 \ & \mathbf{x} \ \in \ \Omega_{i} \\[2mm] 
     -\epsilon_{s} \nabla^{2}\phi_{s}(\mathbf{x})+\epsilon_{s} \kappa_{s}^{2}\phi_{s}(\mathbf{x})=0 & \mathbf{x} \ \in \ \Omega_{s} \\[2mm]
     \phi_{m}(\mathbf{x})= \phi_{i}(\mathbf{x}) & \mathbf{x} \ \in \ \Gamma_{m} \\[2mm]
     \epsilon_{m}\partial_{\mathbf{n}} \phi_{m}(\mathbf{x})= \epsilon_{s}\partial_{\mathbf{n}} \phi_{i}(\mathbf{x}) & \mathbf{x} \ \in \ \Gamma_{m} \\[2mm]
     \phi_{i}(\mathbf{x})= \phi_{s}(\mathbf{x}) & \mathbf{x} \ \in \ \Gamma_{s} \\[2mm]
     \partial_{\mathbf{n}} \phi_{i}(\mathbf{x})= \partial_{\mathbf{n}} \phi_{s}(\mathbf{x}) & \mathbf{x} \ \in \ \Gamma_{s}
\end{matrix}\right. ,
\end{equation}
where for simplicity $\partial_{\mathbf{n}} := \tfrac{\partial}{\partial \mathbf{n}}$ denotes the normal derivative with respect to the interface, and the unit normals on $\Gamma_{m}$ and $\Gamma_{s}$ are oriented as shown in Figure~\ref{fig:2_Iterfaces}. Following 
Figure~\ref{fig:Radio_de_Prueba}, the surface $\Gamma_m$ corresponds to the solvent-excluded surface (SES)~\cite{connolly1983analytical}, while $\Gamma_s$ is located $x$ \AA \ away from the SES, where $x$ is a real number.
 \begin{figure}[h]
 \centering
 \includegraphics[width=0.55\textwidth]{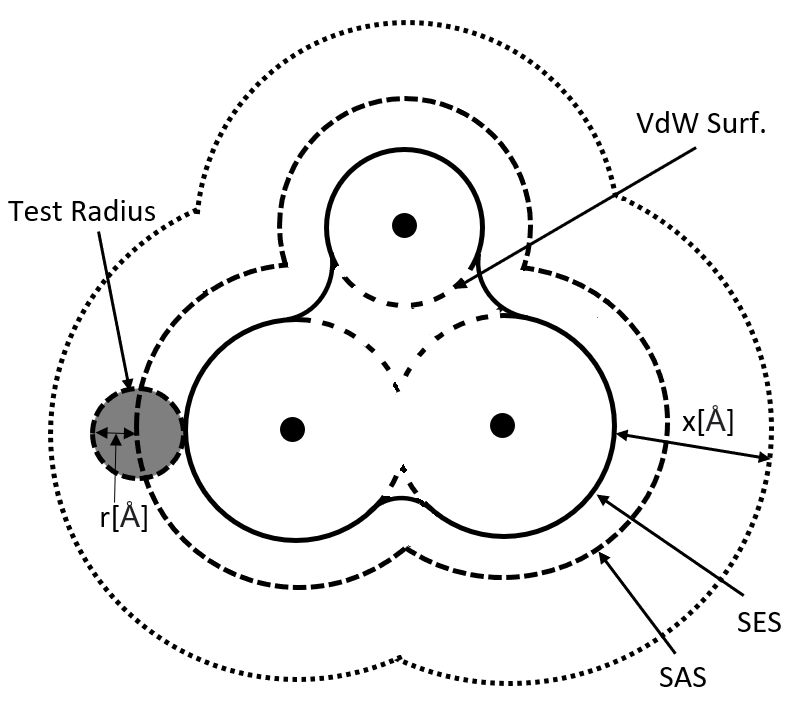}
 \caption{Different surfaces around the molecule. The van der Waals (VdW) and solvent accessible (SAS) surfaces are not used in this work. The SES corresponds to $\Gamma_m$ and the dotted line is $\Gamma_s$. The Test Radius corresponds to the size of one water molecule ($r$=1.4 \AA). }
 \label{fig:Radio_de_Prueba}
 \end{figure}

\subsection{Numerical solution with FEM-BEM coupling}
FEM–BEM coupling is a powerful numerical technique widely used to solve partial differential equations in science and engineering. The Johnson–Nédélec formulation~\cite{johnson1980coupling} is the simplest pairing approach and has been applied to the linear PBE \cite{bsbbbc2023coupling,guerrero2025some}. 

The boundary element method (BEM) does not support the nonlinear form of the equation, nor varying permittivity $\bar{\epsilon}(\mathbf{x})$ or inverse Debye–Hückel length $\bar{\kappa}(\mathbf{x})$. Hence, we use the finite element method (FEM) within $\Omega_{m\cup i} = \Omega_{m} \cup \Omega_{i}$, allowing us to vary $\kappa$ and $\epsilon$ across $\Gamma_m$, and solve the nonlinear PBE in $\Omega_i$. Beyond $\Gamma_s$ we solve the linear PBE with BEM, which naturally adapts to the infinity boundary condition, where the potential decays to zero.


Across the region $\Omega_{m\cup i}$, we are applying integration by parts to the first two equations of the system in Equation~\eqref{eq:pbe_vp} to obtain:
\begin{align*}
\int_{\Omega_{m\cup i}}  \nabla \bar{\epsilon} \phi_{m\cup i} \nabla v\,\mathrm{d}\mathbf{x}&+  \int_{\Omega_{m\cup i}} \bar{\kappa}^{2} \widetilde{N}(\phi_{m\cup i}) \nabla v \,\mathrm{d}\mathbf{x} \\
&-  \int_{\Gamma_s}   \epsilon_s\partial_{\mathbf{n}} \phi_s v\,\mathrm{d}s =  \int_{\Omega_{m\cup i}}\rho_{c} v \,\mathrm{d}\mathbf{x}, 
\end{align*}
where $v \in H_0^1(\Omega_{m\cup i})$ denotes a test function and  
\begin{equation*}
\phi_{m\cup i}(\mathbf{x}) :=
\begin{cases}
\phi_{m}(\mathbf{x}) & \mathbf{x} \in \Omega_m, \\
\phi_{i}(\mathbf{x}) & \mathbf{x} \in \Omega_i.
\end{cases}
\end{equation*}


In the region \(\Omega_{s}\), modelled with BEM, we define the Dirichlet trace operator~\cite{MR2361676} as:
\begin{align*}
\gamma:  H^1(\Omega_s) &\rightarrow H^{\frac{1}{2}}(\Gamma_s), & \gamma f(\mathbf{x}) & := \lim_{\Omega_s \ni \mathbf{y} \rightarrow \mathbf{x} \in \Gamma_s}  f(\mathbf{y}),
\end{align*}

Applying Green's second identity to the third expression of Equation~\eqref{eq:pbe_vp} in \(\Omega_{s}\) \cite{YoonLenhoff1990}, and considering that $\phi_{m\cup i} = \phi_{i}=\phi_{s}$ in $\Gamma_s$, we obtain:
\begin{align*}
\tfrac{1}{2}\gamma\phi_{m\cup i} - K_Y^s\gamma \phi_{m\cup i} + V_Y^s  \lambda_s & = 0.
\end{align*} 
where $\lambda_s := \partial_{\mathbf{n}} \phi_s$, with $\mathbf{n}$ a vector normal to $\Gamma_s$ pointing into $\Omega_s$. In the above equation, $V_Y^s$ and $K_Y^s$ denote the single- and double-layer boundary operators related with the Yukawa kernel (subscript $Y$): 
\begin{align*}
V_Y^s \varphi (\mathbf{x}) &= \int_{\Gamma_s} g_Y(\mathbf{x},\mathbf{x}')\varphi(\mathbf{x}')\mathrm{d}\mathbf{x}',\\
K_Y^s \varphi (\mathbf{x}) &= \int_{\Gamma_s} \frac{\partial g_Y}{\partial\mathbf{n}'}(\mathbf{x},\mathbf{x}')\varphi(\mathbf{x}')\mathrm{d}\mathbf{x}',
\end{align*}
where $\mathbf{x} \in \Gamma_s$ and $\varphi$ refers to an arbitrary function  defined over the boundary $\Gamma_s$. The fundamental solution for the Yukawa operator in free space is given by: 
\begin{align*}
g_Y(\mathbf{x},\mathbf{x}')=\frac{e^{-\kappa|\mathbf{x}-\mathbf{x}'|}}{4\pi|\mathbf{x}-\mathbf{x}'|}
\end{align*}

The initial step in solving the coupled problem involves: \newline
\textit{Find $ \phi_{m\cup i} \in H^1(\Omega_{m\cup i})$ and $\lambda_s \in H^{-\frac{1}{2}}(\Gamma_s)$ such that, for all $v \in H^1(\Omega_1)$ and $\zeta \in H^{-\frac{1}{2}}(\Gamma_s)$}: 

\begin{align*} 
\int_{\Omega_{m\cup i}}  \nabla \bar{\epsilon} \phi_{m\cup i} \nabla v\,\mathrm{d}\mathbf{x}+  \int_{\Omega_{m\cup i}} \bar{\kappa}^{2} N(\phi_{m\cup i}) \nabla v \,\mathrm{d}\mathbf{x} & \\
-  \int_{\Gamma_s}   \epsilon_s\lambda_s v\,\mathrm{d}s &=  \int_{\Omega_{m\cup i}}\rho_{c} v \,\mathrm{d}\mathbf{x},  \\[2mm] \nonumber
  \int_{\Gamma_s}  \left(\tfrac{1}{2} I - K_Y^s\right) \gamma \phi_{m\cup i} \zeta\,\mathrm{d}s + \int_{\Gamma_s} V_Y^s\lambda_s \zeta\,\mathrm{d}s &=0.
\end{align*}

The discrete version of the system above can be written in the following matrix form:
\begin{align} 
    \label{eq:fembem_matrix}
    \begin{bmatrix}
     \widetilde{A} +\bar{\kappa}^{2}M
     & -\epsilon_{s}\widetilde{M
     }^T \\[1mm]
     \left(\frac{1}{2}\widetilde{M
     }-K\right)\gamma & V
    \end{bmatrix} \begin{bmatrix}
    \phi_{m\cup i} \\[1mm] \lambda_{s} 
    \end{bmatrix} =\begin{bmatrix}
    \varrho_{c}\\[1mm] 0 
    \end{bmatrix}.
\end{align}


where the matrices are associated with the scalar products as follows:
\begin{align*}
     \widetilde{A} \sim \int_{\Omega_{m\cup i}}  \nabla \bar{\epsilon} \phi_{m\cup i} \nabla v\,\mathrm{d}\mathbf{x} &,\quad \quad 
 M \sim \int_{\Omega_{m\cup i}} N(\phi) \nabla v \,\mathrm{d}\mathbf{x} & \varrho_{c} \sim \int_{\Gamma_s} \rho_c \zeta\,\mathrm{d}s,\\
 \widetilde{M} \sim \int_{\Gamma_s}  \phi \zeta\,\mathrm{d}s &, \quad \quad K \sim \int_{\Gamma_s}  K_Y^s \phi \zeta\,\mathrm{d}s, & V \sim  \int_{\Gamma_s} V_Y^s \lambda \zeta\,\mathrm{d}s. 
\end{align*}

\color{black}

\subsection{Solvation energy}
The solvation energy corresponds to the free energy associated with dissolving a solute in a polarized solvent. The polar contribution to the total energy can be computed using the following expression \cite{gilson1988calculation}:
\begin{align} 
    \label{eq:Energia}
    E=&\int_{\Omega }\Bigg(\frac{1}{2}\rho(\mathbf{x})\phi(\mathbf{x})+e_{c}I_{0}\phi(\mathbf{x})sinh\left(\frac{e_{c}\phi(\mathbf{x})}{k_BT}\right) &\\ \nonumber
    & -k_{B}T\sum_{n=1}^{2}I_{0}\left(e^{-\frac{e_{c}\phi(\mathbf{x})}{k_BT}}-1 \right) \Bigg) \mathrm{d}\mathbf{x}&
\end{align}  
where $\rho(\mathbf{x})$ is the charge density, \(k_{B}\) is the Boltzmann constant, \(T\) the absolute temperature, \(e_c\) the elementary charge, and $I_{0}$ the bulk ionic strength.

At equilibrium, ions arrange according to Boltzmann statistics, which is a solution of the PBE. Replacing the solution of the PBE in Equation~\eqref{eq:Energia}, and taking the energetic difference between the dissolved and vacuum states yields \cite{gilson1988calculation}:
\begin{equation}
    \label{eq:Esolv}
    \Delta G_{Sol}=\frac{1}{2}\sum_{k=1}^{N_{q}}Q_{k}\phi_{r}(\mathbf{x}_k) + \Delta G_{\Omega_{i}} 
\end{equation}
where $\phi_{r}$ corresponds to the reaction potential (the difference in potential between the solvated and vacuum states). Considering that in vacuum only the Coulomb potential is present, we can write:
\begin{equation*}
    \phi_r =  \phi_m - \phi_{coul} = \phi_m - \frac1{4\pi\epsilon_m}\sum_{k=1}^{N_q}  \frac{Q_k}{|\mathbf{x}_{a} - \mathbf{x}_k|}
\end{equation*}
On the other hand, 
\begin{equation*}
    \Delta G_{\Omega_{i}} =\frac{1}{2}\epsilon_{s}\kappa_{s}^{2}\int_{\Omega_{i} }\Big(\phi_{i}(\mathbf{x})sinh\left(\phi_{i}(\mathbf{x})\right) -2cosh\left(\phi_{i}(\mathbf{x})\right)+2\Big) \mathrm{d}\mathbf{x}. 
\end{equation*}
For the linear PBE, energetic difference in Equation~\eqref{eq:Esolv} reduces to: 
\begin{equation} 
    \label{eq:Esolv_Lineal}
    \Delta G_{Sol}^{L}=\frac{1}{2}\sum_{k=1}^{N_{q}}Q_{k}\phi_{r}(\mathbf{x}_k) 
\end{equation}   


\subsubsection{Calculation of the reaction potential.}


At the solute boundary $\Gamma_m$, the Dirichlet trace for the solute region $\Omega_{m}$ is introduced as: 
\begin{align*}
\overline{\gamma}:  H^1(\Omega_m) &\rightarrow H^{\frac{1}{2}}(\Gamma_m), & \overline{\gamma} f(\mathbf{x}) & := \lim_{\Omega_m \ni \mathbf{y} \rightarrow \mathbf{x} \in \Gamma_m}  f(\mathbf{y}),
\end{align*}

Considering $\epsilon_m$ is constant throughout $\Omega_m$, we can apply Green's second identity on the first expression of Equation~\eqref{eq:pbe_vp} to obtain the boundary integral equation for $\mathbf{x}\in\Omega_m$~\cite{YoonLenhoff1990}:
\begin{equation*}
\phi_{m}(\mathbf{x}) = \mathcal{V}_L^m  \lambda_m(\mathbf{x}) - \mathcal{K}_L^m\phi_{m}(\mathbf{x}) + \phi_{coul}(\mathbf{x})
\end{equation*}
%
being $\lambda_m := \partial_{\mathbf{n}} \phi_m$, with $\mathbf{n}$ a vector normal to $\Gamma_m$ pointing into $\Omega_i$, and
\begin{align}\label{eq:KV}
V_L^m \varphi (\mathbf{x}) &= \int_{\Gamma_m} g_L(\mathbf{x},\mathbf{x}')\varphi(\mathbf{x}')\mathrm{d}\mathbf{x}',\nonumber\\
K_L^m \varphi (\mathbf{x}) &= \int_{\Gamma_m} \frac{\partial g_L}{\partial\mathbf{n}'}(\mathbf{x},\mathbf{x}')\varphi(\mathbf{x}')\mathrm{d}\mathbf{x}',
\end{align}
where $\mathcal{V}_L^m$ and $\mathcal{K}_L^m$ denote the single-layer and double-layer boundary operators related with the Laplace kernel (subscript $L$), $\varphi(\mathbf{x})$ represents any surface function defined on $\Gamma_m$, and the Laplace fundamental solution in free space is: 
\begin{align*}
g_L(\mathbf{x},\mathbf{x}')=\frac{1}{4\pi|\mathbf{x}-\mathbf{x}'|}. 
\end{align*}
The reaction potential can then be computed as
\begin{equation}
\label{eq:Potential_in_Omega_m}
\phi_r = \phi_{m} - \phi_{coul} = \mathcal{V}_L^m  \lambda_m - \mathcal{K}_L^m\overline{\gamma}\phi_{m}.
\end{equation}

After solving Equation~\eqref{eq:fembem_matrix}, we obtain $\overline{\gamma}\phi_{m}$ by interpolating between the volumetric and surface meshes of $\Omega_{m\cup i}$ and $\Gamma_m$, respectively. However, $\lambda_m$ requires solving an extra boundary integral equation. Taking the limit $\mathbf{x} \to\Gamma_m$ in~\eqref{eq:Potential_in_Omega_m}, we obtain:
\begin{align*}
\tfrac{1}{2}\overline{\gamma}\phi_{m}(\mathbf{x}) + K_L^m\overline{\gamma} \phi_{m}(\mathbf{x}) - V_L^m  \lambda_m(\mathbf{x}) & =  \phi_{coul}(\mathbf{x})
\end{align*}
Here, $V_L^m$ and $K_L^m$ denote the single-layer and double-layer boundary operators as Equation~\eqref{eq:KV}, but in this case, $\mathbf{x} \in\Gamma_m$. 
We can then obtain $\lambda_m$ from:
\begin{align*}
V_L^m  \lambda_m(\mathbf{x}) & = \tfrac{1}{2}\overline{\gamma}\phi_{m}(\mathbf{x}) + K_L^m\overline{\gamma} \phi_{m}(\mathbf{x}) - \phi_{coul}(\mathbf{x}).
\end{align*}

\subsection{Solving the Nonlinear Equation.}
To solve the nonlinear FEM–BEM system, we first define  \(\Phi_{T}\) as a column vector containing $[\phi, \lambda]$ as in Equation~\eqref{eq:fembem_matrix}. We can decompose it in linear and nonlinear contributions as:
\begin{equation}
    \label{eq:Ec_phi_T}
    \Phi_{T}=\Phi_{L}+\Phi_{NL} 
\end{equation} 

The linear component \(\Phi_{L}\) is computed by directly solving the matrix system from Equation~\eqref{eq:fembem_matrix} using $N(\phi(\mathbf{x})) = \phi(\mathbf{x})$. 

Using $N(\phi(\mathbf{x})) = sinh(\phi(\mathbf{x}))$ we obtain the nonlinear FEM–BEM system as given in Equation~\eqref{eq:fembem_matrix}. Since nonlinearity is only present in the intermediate domain $\Omega_i$ we can split our system for two matrices:
\begin{align}
    \Ma=\begin{bmatrix}
     \widetilde{A} 
     & -\epsilon_{s}\widetilde{M}^T \\[1mm]
     \left(\frac{1}{2}\widetilde{M}-K\right)\gamma & V
    \end{bmatrix} &,&  
    \Vcc=\begin{bmatrix}
     \bar{\kappa}^{2}M 
     & 0 \\[1mm]
     0 & 0
    \end{bmatrix}. 
\end{align}
Despite that the nonlinear terms correspond to the whole FEM volume domain, we recall that according to the second term in Equation~\eqref{eq:Variables_2Superficie}, operator $M$ will be zero on the whole domain $\Omega_m$.

The solution of the nonlinear PBE requires solving:
\begin{equation} 
    \label{eq:general_nonlinear}
    \Ma\Phi_{T} + \Vcc\left(\Phi_{T}\right)=\Vb,
\end{equation} 
where
\begin{equation}
    \label{eq:Vector_c}
    \Vcc(\Phi) = 
    \begin{bmatrix}
        \bar{\kappa}^{2}M \phi\\[1mm] 0 
    \end{bmatrix} =
    \begin{bmatrix}
        \bar{\kappa}^{2}\int_{\Omega_{m\cup i}} sinh(\phi) \nabla v \,\mathrm{d}\mathbf{x} \\[1mm] 0 
    \end{bmatrix}.
\end{equation}

Below, we explore two iterative solvers based on Picard and Newton-Raphson methods.

\subsubsection{Picard method.} 

Picard method treats the nonlinear term using the previous iterate, hence our linearised system of Equation~\eqref{eq:general_nonlinear} becomes:
\begin{equation*} 
    \Ma\Phi_{T}^{k+1}=\Vb-\Vcc(\Phi_{T}^{k})
\end{equation*} 
Substituting the decomposition from Equation~\eqref{eq:Ec_phi_T} into the expression above, we obtain:
\begin{equation}  
    \label{eq:Ec_NL_Previo2} 
    \Ma\Phi_{NL}^{k+1}=-\Ma\Phi_{L}+\Vb-\Vc \quad \quad k=0,1,2...
\end{equation} 
However, using the estimate ${\Phi}_{NL}^{k+1}$ directly in the next iteration $(k+2)$ often leads to divergence of the iterative method. Thus, instead of ${\Phi}_{NL}^{k+1}$, we introduced $\widetilde{\Phi}_{NL}^{k+1}$ as an intermediate estimate of the nonlinear potential at iteration $k+1$, obtained by solving the Equation~\eqref{eq:Ec_NL_Previo2},
and further updating $\Phi_{NL}^{k+1}$ using a relaxation factor $\omega_k\in[0,1]$ such that: 
\begin{equation}
\label{eq:Ec_NL_Previo3}
\Phi_{NL}^{k+1}=\omega_{k+1}\widetilde{\Phi}_{NL}^{k+1}+(1-\omega_{k+1})\Phi_{NL}^{k}
\end{equation} 

The Picard iteration is terminated once the norm of the difference between successive iterates falls below a predefined tolerance:
\begin{equation}
       \label{eq:Norma_Picard}
       e_{k+1}=\left\|\Phi_{NL}^{k+1}-\Phi_{NL}^{k} \right\|_{2}= \omega_{k+1}\left\|\widetilde{\Phi}_{NL}^{k+1}-\Phi_{NL}^{k} \right\|_{2}<\textbf{Tol}_{NL}
\end{equation} 

\subsubsection{Newton-Raphson method.} 
The Newton–Raphson method solves the nonlinear system using the Jacobian matrix. 
Applying Taylor's theorem to Equation~\eqref{eq:general_nonlinear} results in the following Newton–Raphson formulation:
\begin{equation} 
\label{eq:NR_main}
\left(\Ma + \Vdcf\right)\Delta\Phi_{NL}^{k+1}=-\Ma\Phi_{L}-\Ma\Phi_{NL}^{k}+\Vb-\Vc
\end{equation}

Once again, we introduce 
%
a relaxation factor $\omega_{k+1}$ to help convergence of the Newton–Raphson iteration. The updated nonlinear potential is computed as: 
\begin{equation}       
    \label{eq:Ec_NL_NR}
    \Phi_{NL}^{k+1}=\Phi_{NL}^{k}+\omega_{k+1}\Delta\Phi_{NL}^{k+1} 
\end{equation} 
Similarly to Picard, the Newton–Raphson iteration is terminated once the change in the solution falls below a prescribed tolerance, measured in the norm:
\begin{equation}
       \label{eq:Norma_Newton-Raphson}
       e_{k+1}=\left\|\Phi_{NL}^{k+1}-\Phi_{NL}^{k} \right\|_{2}= \omega_{k+1}\left\|\Delta{\Phi}_{NL}^{k+1} \right\|_{2}<\textbf{Tol}_{NL}
\end{equation} 

\subsection{Calculation of the optimal relaxation factor.}

In our analysis, we are allowing the relaxation factor $\omega_{k}$ to vary throughout iterations. 
Selecting a good relaxation factor is critical to achieve optimal convergence of the nonlinear system. Manual selection of \(\omega_{k}\) is not a reliable strategy, as it can lead to slow convergence 
and requires an inefficient interaction with the user. This challenge is further complicated considering the optimal \(\omega_{k}\) ($\omega^{opt}_{k}$) can vary significantly depending on the solute geometry and charge. 
We address this issue minimising the norm in Equation~\eqref{eq:Norma_Picard} for Picard, and using the global convergence condition for Newton-Raphson, to compute $\omega^{opt}_{k}$ at each iteration $k$. To the best of the authors’ knowledge, this is the first automated strategy for selecting the relaxation parameter in the Picard iterations for the nonlinear PBE. Previously, Holst and Saied \cite{holst1995numerical} proposed an automatic process to determine optimal relaxation factors within an inexact Newton framework; however, their method is considerably more complex. In contrast, our approach for both the Picard and Newton-Raphson methods is simpler and easier to implement.


\subsubsection{Picard method}
To compute $\omega^{opt}_{k}$ for the Picard method, we start from the expression for the error \(e_{k+1}\) from Equation~\eqref{eq:Norma_Picard} 
and using Cauchy-Schwarz inequality to obtain: 
\begin{align*}
    e_{k+1}(\omega_{k})  &= \left\|\Phi_{NL}^{k+1}(\omega_{k+1})-\Phi_{NL}^{k}(\omega_{k}) \right\|_{2}\frac{\left\|\Ma \right\|_{2}}{\left\|\Ma \right\|_{2}}   \\ 
    &\ge \frac{\omega_{k+1}\left\| \Ma\left(\widetilde{\Phi}_{NL}^{k+1}(\omega_{k})-\Phi_{NL}^{k}(\omega_{k})\right) \right\|_{2}}{\left\|\Ma \right\|_{2}}
    \\
    & = \frac{\omega_{k+1}\left\| \zd \right\|_{2}}{\left\|\Ma \right\|_{2}} 
     = \widetilde{e}_{k+1}(\omega_{k}).    
\end{align*}
Let us note that from Equation~\eqref{eq:Ec_NL_Previo2}, it follows
\begin{equation*}
    \zd= -\Ma\Phi_{L}-\Ma\Phi_{NL}^{k}(\omega_{k})+\Vb-\Vcc\left(\Phi_{L}+\Phi_{NL}^{k}(\omega_{k})\right)
\end{equation*}



Since $\Ma$ is independent of $\omega_{k}$, we minimise $\widetilde{e}_{k+1}(\omega_{k})$ by computing the first derivative of the square (for simplicity) 
and setting it to zero:
\begin{equation}
\label{eq:Picard_optimise_omega}
       \frac{\partial \widetilde{e}_{k+1}^2(\omega_{k})}{\partial \omega_{k}} =\frac{2\omega_{k+1}^2\zd\cdot\zdd}{\left\|\Ma \right\|_{2}^2}=0 .
\end{equation} 

The only viable condition to achieve a minimum is to nullify the scalar product in the numerator. 

\subsubsection{Newton-Raphson method}
To ensure optimal convergence for the Newton-Raphson method, we use the global convergence condition, as in the work of Holst and Saied~\cite{holst1995numerical}. The condition requires the next step to be closer to zero than the previous one. In our notation it can be expressed as follows:
\begin{align*}
\|-\Ma\Phi_{L}-\Ma(\Phi_{NL}^{k-1}+\omega_{k}\Delta{\Phi}_{NL}^{k})+\Vb-\Vcc(\Phi_{L}+\Phi_{NL}^{k-1}+\omega_{k}\Delta{\Phi}_{NL}^{k})\|_2 \\ < \|-\Ma\Phi_{L}-\Ma\Phi_{NL}^{k-1}+\Vb-\Vcc(\Phi_{L}+\Phi_{NL}^{k-1})\|_2 
\end{align*}
 Let us notice that it is equivalent to:
\begin{equation*}
    \|\mathbf{d}(\Phi_{NL}^{k}(\omega_{k}))\|_2 < \|\mathbf{d}(\Phi_{NL}^{k-1}(\omega_{k-1}))\|_2.
\end{equation*}
To compute $\omega^{opt}_{k}$ and fulfil the above condition, we choose to minimise $\|\zd\|_2^2$ using derivative with respect to $\omega_k$
\begin{equation}
\label{eq:Newton_optimise_omega}
       \frac{\partial \zd
       \cdot \zd}{\partial \omega_{k}} =2\zd\cdot\zdd=0. 
\end{equation} 
Once again, the only viable condition 
is to nullify the scalar product.

\subsubsection{Optimal relaxation factor}
Despite different arguments for each method, the conditions in Equations~\eqref{eq:Picard_optimise_omega} and~\eqref{eq:Newton_optimise_omega} are very similar. For convenience, we define the auxiliary function \(g_d\) representing:  
\begin{equation}
    \label{eq:gd=0}
    g_d(\omega_{k})=\zd\cdot\zdd = 0.
\end{equation} 
which we can express using auxiliary variables $\za$, $\zb$ and $\zdb$ giving
\begin{equation*}
    g_d(\omega_{k})=(\za \omega_{k}+\zb)\cdot (\za+\zdb) = 0.
\end{equation*} 
%
The auxiliary terms are defined as:
\begin{align}
\label{eq:ab_pic}\nonumber
    \za &= -\Ma \zc
    \\
    \zb &= -\Ma\Phi_{L}-\Ma\Phi_{NL}^{k-1}+\Vb-\Vcc\left(\Phi_{L}+\Phi_{NL}^{k}(\omega_{k})\right) \\ \nonumber
    \zdb  &= - \zc \partial_{\Phi}\Vcc\left(\Phi_{L}+\Phi_{NL}^{k}(\omega_{k})\right) 
\end{align}
where
\begin{equation*}
    \zc = 
    \left\{\begin{matrix} 
    \widetilde{\Phi}_{NL}^{k}-\Phi_{NL}^{k-1} &
    \textbf{Picard} \\ 
    \Delta{\Phi}_{NL}^{k} & \textbf{Newton-Raphson}
    \end{matrix}\right. .
\end{equation*}

By performing algebraic manipulations to isolate \(\omega_{k}\), we arrive at the following form of the auxiliary function \(g_d\) in terms of a new scalar function \(f_d\):
\begin{equation}
    \label{eq:fd=0}  
    f_d(\omega_{k})=-\frac{(\za+\zdb)\cdot\zb}{(\za+\zdb)\cdot\za}-\omega_{k} =0
\end{equation} 
The function $f_d(\omega_{k})$ makes sense when denominator is non zero, and we have experimentally confirmed that it is a case.

In the first section of the results (Figures~\ref{fig:gd_esfera} and \ref{fig:fd_esfera}), it is shown that working with \(f_d\) is generally more practical than using \(g_d\). This is because \(g_d\) 
can become very large, especially during early iterations. 
In contrast, \(f_d\) evaluates to much smaller values, since it involves ratios of dot products, which facilitates more efficient and stable computation of the optimal \(\omega_{k}\). 

We call the optimal relaxation factor for the $k$-th iteration ($\omega_k^{opt}$) the value of $\omega_k$ such that the conditions in Equations~\eqref{eq:Picard_optimise_omega} or~\eqref{eq:Newton_optimise_omega}, and hence Equation \eqref{eq:fd=0}, are met. We used three different iterative methods to find the root of Equation~\eqref{eq:fd=0} and obtain $\omega^{opt}_{k}$: Bisection, Newton–Raphson, and the Secant method, and combinations between them. Their implementations are described in \ref{sec:one_dim_root_finding}.

%% file: Resultados_Principal.tex
\section{Results}



\begin{figure}
    \centering
    \includegraphics[width=0.7\textwidth]{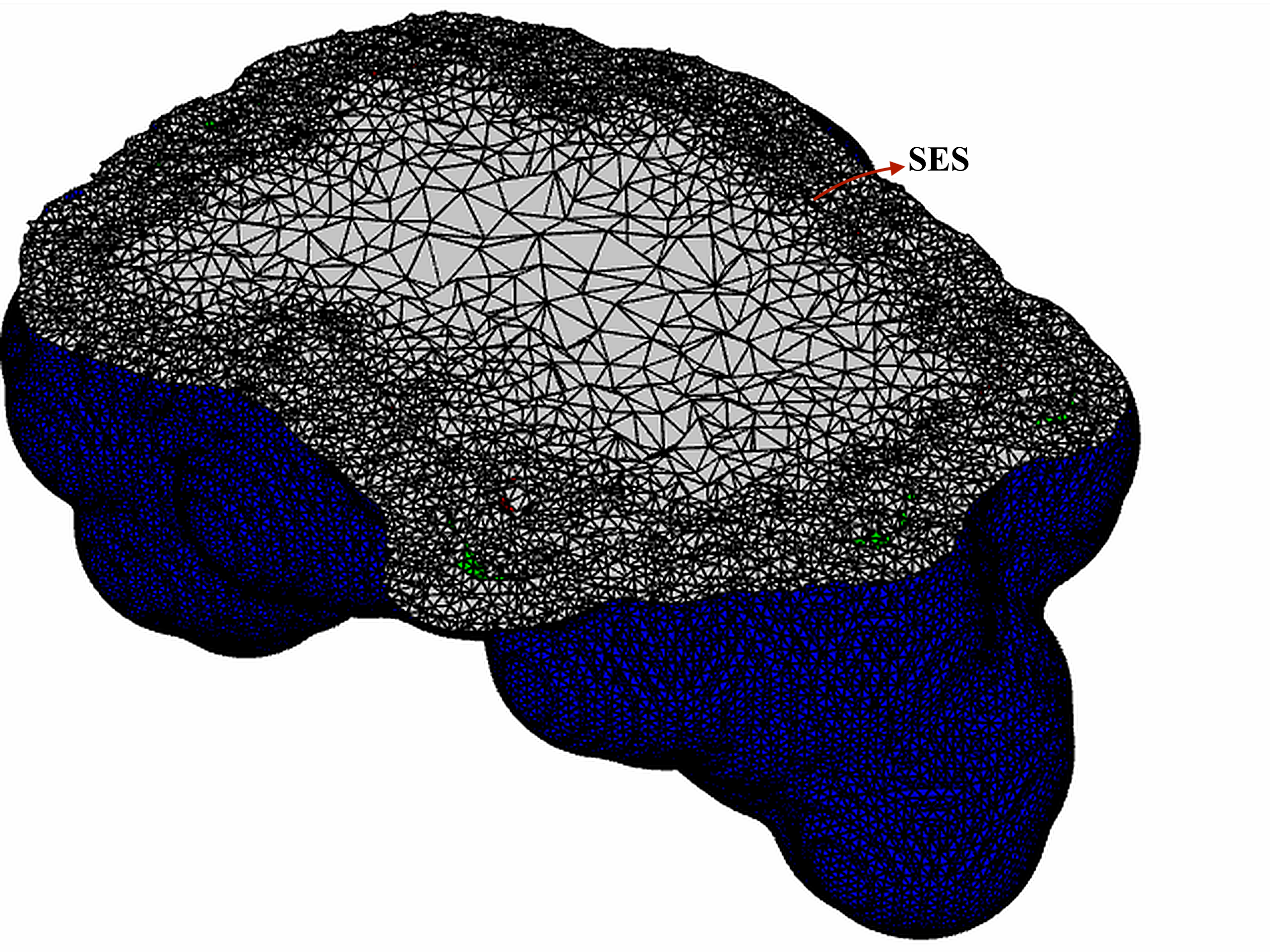}
    \caption{Cross section of an example FEM mesh (in grey), with the SES identified within (fine mesh region close to the interface). The BEM mesh (equivalent to the outer boundary of the FEM mesh) is identified in blue.}
    \label{fig:fem_bem_mesh}
\end{figure}


The results section begins with a validation of our nonlinear FEM–BEM solver against the reference implementation provided by APBS \cite{BakerETal2001,JurrusETal2018} for a spherical test case. We then compute the polar component of the solvation free energy of the PDB structure 1AJF to evaluate different algorithmic approaches. Owing to its moderate size and high charge density, the RNA structure of 1AJF provides a representative benchmark for analysis optimal convergence behaviour for various solvers in highly charged systems. Finally, the optimal configuration is tested on five highly charged molecules to evaluate the robustness and generalization of the approach.

All simulations were performed on a workstation equipped with two Intel(R) Xeon(R) E5-2680 v3 CPUs (2.50 GHz, 12 cores each) and 96 GB of RAM.
The computational framework combines \texttt{Bempp-cl} 0.3.1 \cite{betcke2021bempp} and the \texttt{Dolfin} interface from \texttt{FEniCS} 2019.1.0 \cite{logg2010dolfin}, running under \texttt{Python} 3.8.
The surface and volume meshes were generated and processed using \texttt{Trimesh} \cite{trimesh}, \texttt{MSMS} \cite{sanner1996reduced}, \texttt{Nanoshaper} \cite{decherchi2013general}, \texttt{PyGAMER} \cite{lee2020open}, and \texttt{PBJ} \cite{search2022towards}.
All input files and scripts required to reproduce the results in this section are publicly available in Zenodo\footnote{\url{https://zenodo.org/records/17633060}} and Github\footnote{\url{https://github.com/bem4solvation/Nonlinear_FEM-BEM_Coupling}} repositories, respectively. 

The finite-element tetrahedral meshes were generated from triangular surface meshes defined on $\Gamma_m$ (the solvent-excluded surface, SES) and $\Gamma_s$.
We used the surface vertex density to control the overall grid resolution.
Figure \ref{fig:fem_bem_mesh} is an example of the resulting discretisation, where the SES is clearly visible within the FEM domain.

For relatively small systems, the boundary operators were assembled densely, with \texttt{Bempp-cl} explicitly forming the BEM matrices.
For larger systems (surface meshes that exceed 90,000 vertices), we employed the Fast Multipole Method (FMM) \cite{wang2021high} to accelerate matrix–vector products.
Both the BEM and FEM formulations used piecewise-linear (P1) basis functions, with fourth-order Gaussian quadrature for numerical integration.

The result tables report the electrostatic solvation energy (ESE), decomposed into its linear component (LC), obtained from Equation~\eqref{eq:fembem_matrix} with $N(\phi) = \phi$, and its nonlinear component (NC), computed using Equation~\eqref{eq:Ec_NL_Previo3} for the Picard method or Equation~\eqref{eq:Ec_NL_NR} for the Newton–Raphson approach. Additionally, the tables list the solvation free energy contribution $\Delta G_{\Omega_{i}}$ from Equation~\eqref{eq:Esolv}. Performance columns include the total runtime, preprocessing time (PT), time spent solving the nonlinear iterative system (NIS), and the number of nonlinear iterations (I).

\subsection{Validation against APBS on a spherical domain}

We validated our nonlinear PBE solver by comparing to APBS on energy calculations for a sphere. The spherical geometry had an inner radius of 2 \AA~(for $\Gamma_m$), an outer radius of 5 \AA~(for $\Gamma_s$), and contained three point charges of magnitude $q$=1$e$ located at (0,0,0) and (0,$\pm$0.5,0) \AA. Surface meshes were generated using \texttt{MSMS} for different density levels according to Table \ref{table:Informacion_1Capa}. The internal volumetric mesh was generated using \texttt{Tetmesh} with a growth factor of 1.1 applied to adjacent tetrahedra (limiting element growth to less than 10\%). Table \ref{table:Informacion_1Capa} summarises the number of vertices and elements for both the surface and volumetric meshes corresponding to each configuration.


\begin{table}[ht]
\centering
\footnotesize
\caption{Vertex and element information for the surface and volumetric meshes for the sphere.}
\begin{tabular}{|c|c|c|c|c|c|c|} 
\hline
& \textbf{MSMS} & \multicolumn{4}{c|}{\textbf{Surface Mesh}} & \textbf{Vol.}\\
\cline{3-6}
& \textbf{Dens.}  & \multicolumn{2}{c|}{\textbf{Radius 2 \AA~}} & \multicolumn{2}{c|}{\textbf{Radius 5 \AA~}} & \textbf{Mesh}\\
\cline{3-6}
\textbf{Case}  & \textbf{vert/\AA$^2$}  & \textbf{Vert.} &  \textbf{Elem.} & \textbf{Vert.} & \textbf{Elem.} & \textbf{Vert.} \\ 
\hline
0 & 1 & 46 & 88	& 290 & 576	 & 449 \\
1 & 1.9	& 85 & 166 & 566 & 1128	 & 1026 \\
2 & 3.8	& 180 & 356	& 1149 & 2294  & 2312 \\
3 & 7.5	& 360 & 716	& 2334 & 4664  & 5076 \\
\hline
\end{tabular}
\label{table:Informacion_1Capa}
\end{table}


All cases assume a solute with permittivity $\epsilon_{m}=4$, and model the solvent with a relative permittivity of $\epsilon_{s}=80$ and an inverse Debye length of $\kappa$=0.125 \AA$^{-1 }$, corresponding to a 150 mM NaCl aqueous solution. 

\begin{figure}[h]
\centering
\includegraphics[width=0.8\textwidth]{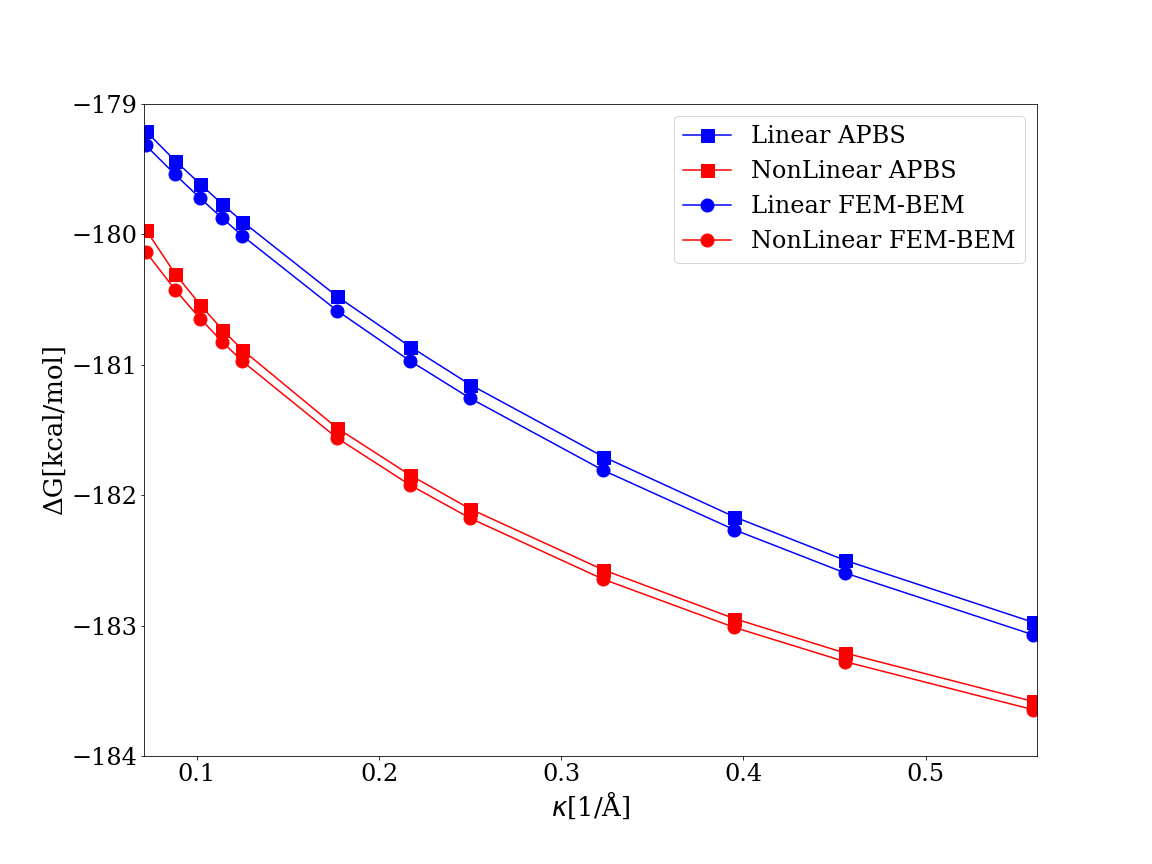}
\caption{Solvation energy of a sphere for different values of $\kappa$ in linear and nonlinear FEM-BEM compared to APBS.}
\label{fig:Graf_EvsK}
\end{figure}

To solve the nonlinear PBE we used Picard iterations with a tolerance of $Tol_{NL}=10^{-3}$ and a constant relaxation factor $\omega^{con}=0.16$. We set the GMRES solver to a fixed tolerance of $Tol_{GR}=10^{-6}$. The computed results are compared with those obtained using APBS for various values of $\kappa$, using grid dimensions of 65, 129 and 257, on a 34$\times$34$\times$34 \AA~computational box. We used the three APBS refinements to compute a Richardson extrapolated value of the energy, which we compared against the extrapolation obtained with the three finest meshes from Table \ref{table:Informacion_1Capa}, which are plotted in Figure \ref{fig:Graf_EvsK}.


Figure~\ref{fig:Graf_EvsK} shows that for varying values of $\kappa$ in the spherical model, the energy results obtained using our method are in good agreement with those computed by APBS. This serves as verification of the correct solution of the nonlinear Poisson–Boltzmann equation.


\subsection{Behaviour of $f_d$ and $g_d$.} 

The optimal relaxation factor $\omega^{opt}_k$ can be obtained from the function $f_d$ in Equation~\eqref{eq:fd=0} or the function $g_d$ in Equation~\eqref{eq:gd=0}. The motivation of this section is to examine the behaviour of both auxiliary functions, and choose which is most efficient to find $\omega^{opt}_k$. 

We performed multiple simulations of Case 2 of the spherical model described in Table \ref{table:Informacion_1Capa}, solving the nonlinear Poisson–Boltzmann equation with both Picard and Newton–Raphson methods with a tolerance of $Tol_{NL}=10^{-3}$. We introduced the sinh function gradually using its cubic-order Taylor expansion in the initial iteration only. The $\omega^{opt}_k$ to advance in iterations was computed using Newton–Raphson method with a tolerance of $Tol_\omega=10^{-2}$ and the GMRES solver was set to a tolerance of $Tol_{GR}=10^{-6}$. 


\begin{figure}
     \centering
     \begin{subfigure}
         \centering
         \includegraphics[width=0.8\textwidth]{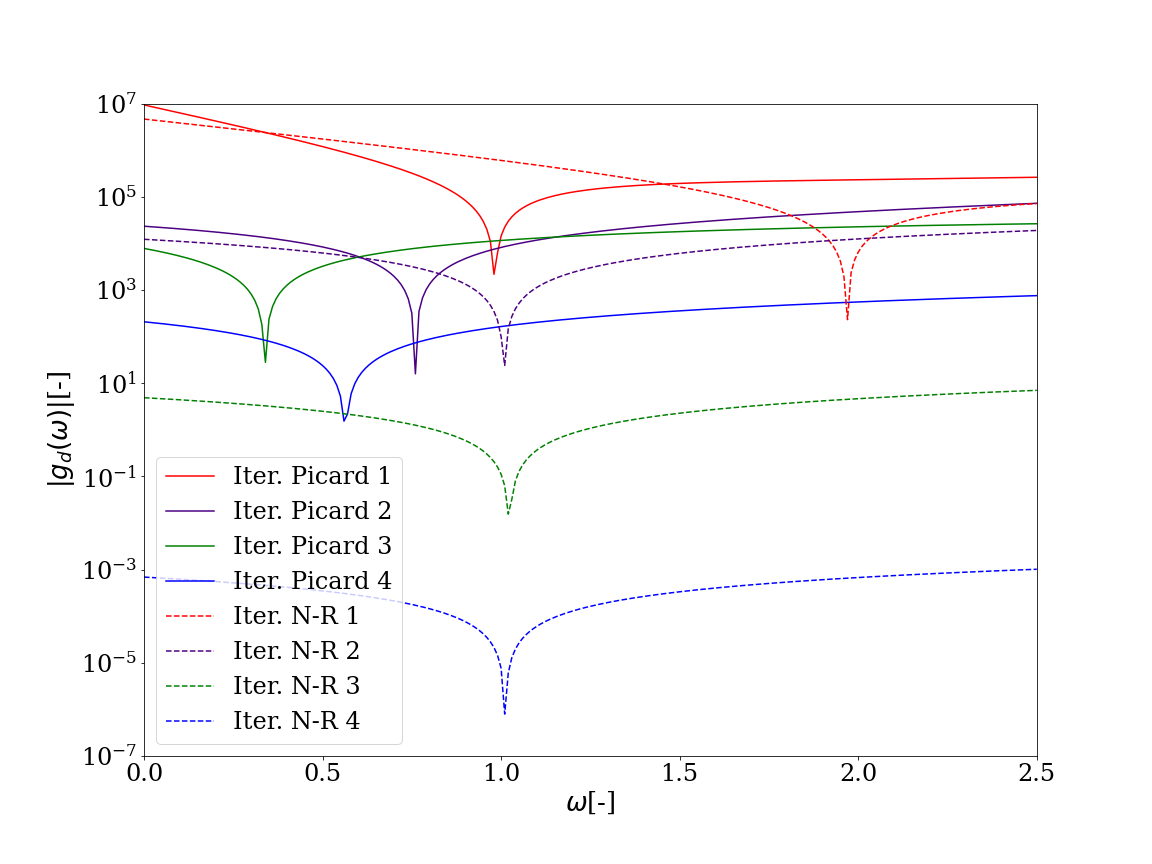}
         \caption{Function $|g_d|$ for the first 4 iterations of Picard (solid lines) and Newton-Raphson (dashed lines) methods on the sphere. Note this is a semilog plot of the absolute value, hence, zeros are located at the cusps, where $g_d$ changes sign.}
         \label{fig:gd_esfera}
     \end{subfigure}
     \begin{subfigure}
         \centering
         \includegraphics[width=0.8\textwidth]{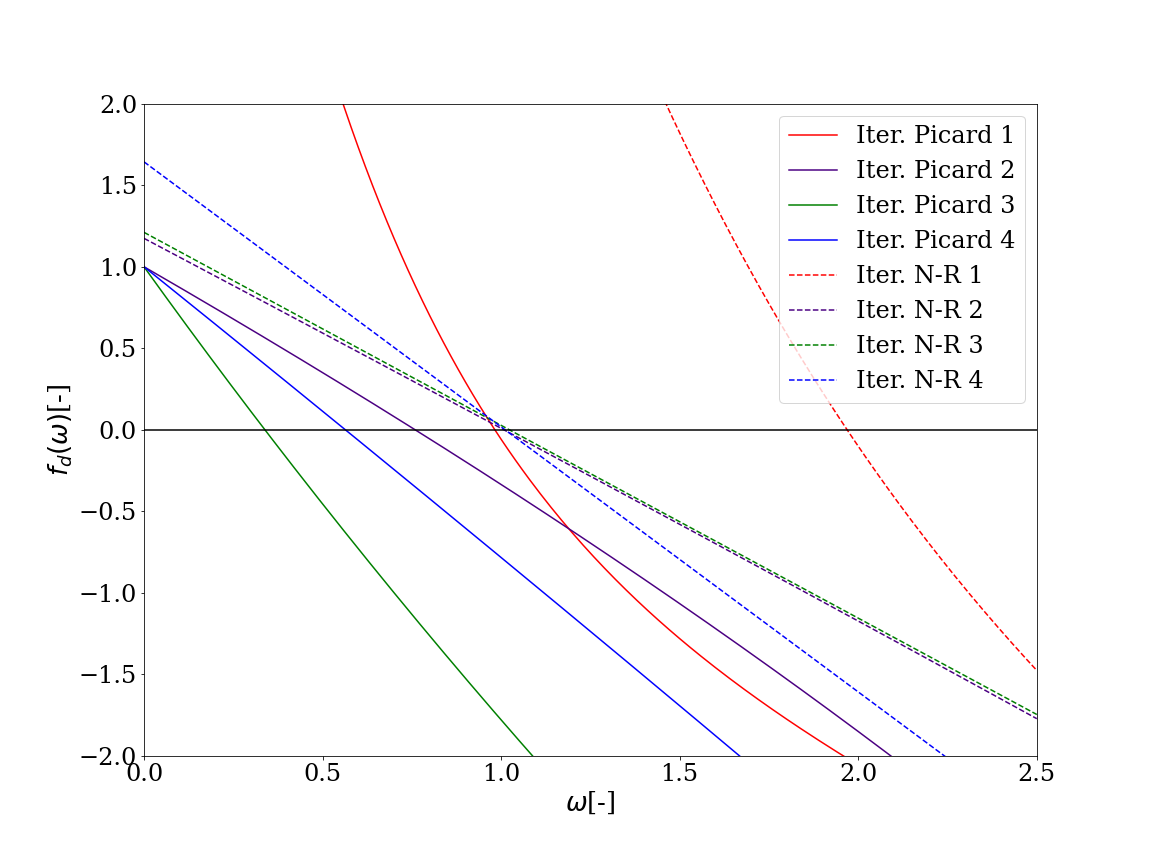}
         \caption{Function $f_d$ for the first 4 iterations of Picard (solid lines) and Newton-Raphson (dashed lines) methods on the sphere.}
         \label{fig:fd_esfera}
     \end{subfigure}
        
\end{figure}


     
Figures~\ref{fig:gd_esfera} and~\ref{fig:fd_esfera} plot $g_d(\omega)$ and $f_d(\omega)$ for the first four Picard and Newton-Raphson iterations. 
The values of $g_d$ are much larger than those of $f_d$, which makes it harder to accurately compute the roots of $g_d(\omega)$ numerically.  Except for the first iteration (performed with a cubic expansion), the relation between $f_d$ and $\omega$ is linear for both methods, and $f_d(0)=1$ consistently for Picard, making $f_d$ more convenient to obtain $\omega^{opt}_{k}$.



\subsection{Algorithm optimisation with 1AJF}

We used 1AJF, an RNA structure that consists of 581 atoms with total charge of -17 $[e]$~\cite{kieft1997solution}, as our model problem to find an optimal algorithm to solve the nonlinear PBE avoiding manual $\omega$ selection. We explored factors like the nonlinear scheme to compute the Poisson-Boltzmann equation and approximate $\Vcc$, GMRES tolerances, and alternatives to compute $\omega_k^{opt}$.

After obtaining the 1AJF structure from the PDB, we parametrised it with AMBER \cite{ponder2003force} using PDB2PQR \cite{dolinsky2004pdb2pqr}. Throughout this section, the \texttt{Nanoshaper}-generated SES ($\Gamma_m$) has 2 vertices/\AA$^2$ (4764 vertices, 9524 elements), as well as $\Gamma_s$ (6926 vertices, 13848 elements), which was placed 3 \AA~away. \texttt{Tetmesh} generated the internal volumetric mesh (25284 vertices) using a 1.1 growth factor ($<$ 10\% growth).

\subsubsection{Gradually incorporating sinh to the nonlinear solver 
}


To avoid divergence, we approximate the nonlinear vector $\Vcc$ (Equation~\eqref{eq:Vector_c}) using Taylor expansions in schemes that increase in order as iterations advance. Schemes are detailed in Tables \ref{table:Picard_Vec_C} and \ref{table:Newton-Raphson_Vec_C}, and are defined by:
\begin{itemize}
    \item \textbf{T3, T5, and T7:} Single iterations using third-, fifth-, and seventh-order Taylor expansions of $\sinh(x)$, respectively.
    \item \textbf{HS:} Full hyperbolic sine function in all subsequent iterations.
\end{itemize}

In all these calculations, the nonlinear solver and GMRES tolerances were set to $Tol_{NL}=10^{-3}$ and $Tol_{GR}=10^{-6}$, respectively.


\begin{table}[ht] 
\centering
\footnotesize
\caption{ESE of 1AJF computed with Picard iterations using constant relaxation factor $\omega^{con}$, for different schemes of the vector $\Vcc$ representing nonlinear component.}
\begin{tabular}{|c|c|c|c|c|c|c|c|c|c|} 
\hline
 & \multicolumn{4}{c|}{\textbf{ESE [kcal/mol]}} & \multicolumn{3}{c|}{\textbf{Time [s]}} & & \\
\cline{2-8}
\textbf{Scheme} & \textbf{Total} & \textbf{LC} & \textbf{NC} & \textbf{$\Delta G_{\Omega_{i}}$} & \textbf{Total}  & \textbf{PT} & \textbf{NIS} & \textbf{I} & \textbf{$\omega^{con}$}  \\
\hline
HS  & -1129.003 & -1122.647 &  -9.477	& 3.121	&  170.36 & 60.73 & 105.07 & 18 & 0.22\\
T3-HS  & -1128.999 & -1122.647 &  -9.474	& 3.121	&  162.74 & 60.83 & 97.15	& 16 & 0.21 \\
T3-T5-HS & -1129.001	& -1122.647 & -9.475	& 3.121	& 163.57 & 60.53 & 98.89 & 17 & 0.21 \\
T3-T5-T7-T7-HS & -1129.001	& -1122.647	& -9.475 & 3.121 & 163.26 & 60.97 & 98.63 & 17 & 0.21 \\
\hline
\end{tabular}
\label{table:Picard_Vec_C}
\end{table}


\begin{table}[ht] 
\centering
\footnotesize
\caption{ESE of 1AJF computed with Newton–Raphson iterations using constant relaxation factor $\omega^{con}$, for different schemes of the vector $\Vcc$ representing nonlinear component.}
\begin{tabular}{|c|c|c|c|c|c|c|c|c|c|} 
\hline
 & \multicolumn{4}{c|}{\textbf{ESE [kcal/mol]}} & \multicolumn{3}{c|}{\textbf{Time [s]}} & & \\
\cline{2-8}
\textbf{Scheme} & \textbf{Total} & \textbf{LC} & \textbf{NC} & \textbf{$\Delta G_{\Omega_{i}}$} & \textbf{Total}  & \textbf{PT} & \textbf{NIS} & \textbf{I} & \textbf{$\omega^{con}$}  \\
\hline
HS & -1129.004	& -1122.647	& -9.478 & 3.121 & 125.54 & 64.77 & 56.98 & 9 & 1.10 \\
T3-HS & -1129.004 & -1122.647 & -9.478 & 3.121 & 109.21 & 64.14 & 41.12	& 7 & 1.10 \\
T3-T5-HS & -1129.004 & -1122.647 & -9.478 & 3.121 & 103.64 & 64.31 & 35.41 & 6 & 1.10 \\
T3-T5-T7-T7-HS & -1129.004 & -1122.647 & -9.478 & 3.121 & 111.39 & 65.00 & 42.85 & 7 & 1.05 \\
\hline
\end{tabular}
\label{table:Newton-Raphson_Vec_C}
\end{table}

\begin{figure}
    \centering
    \includegraphics[width=0.7\textwidth]{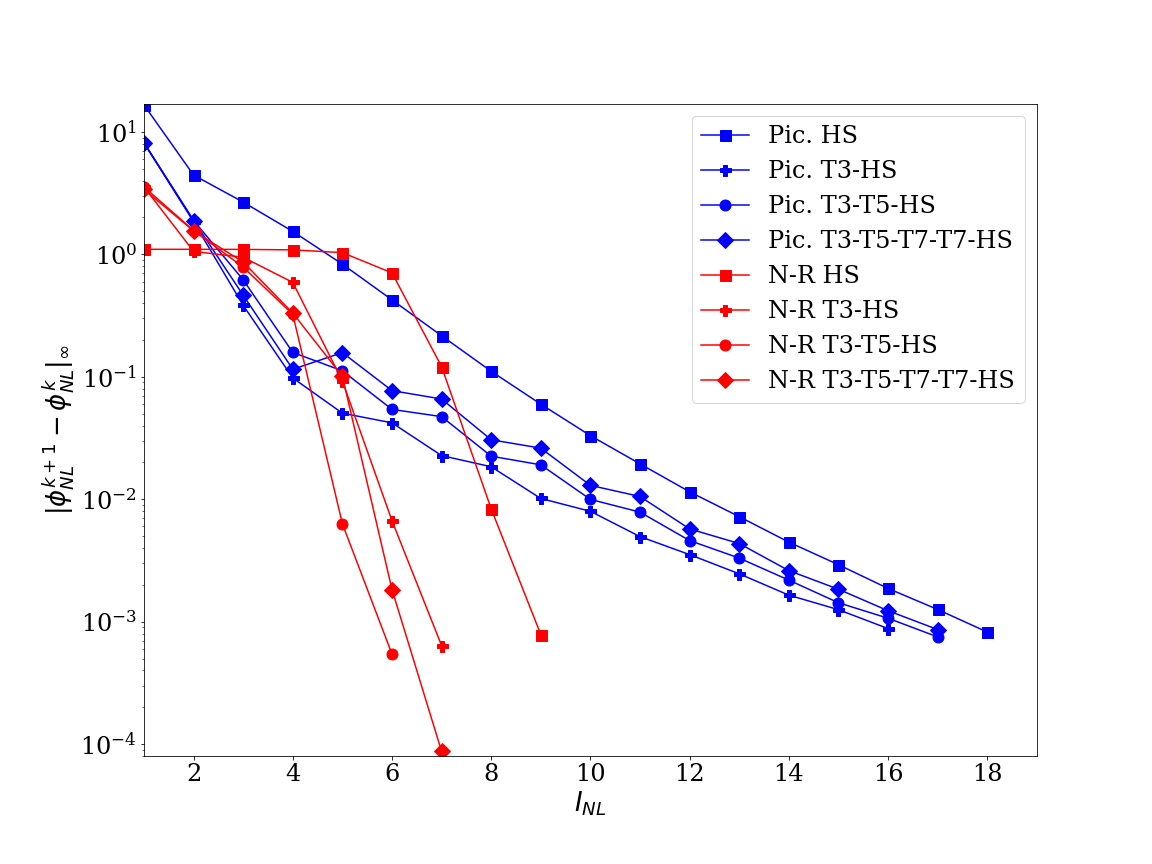}
    \caption{Evolution of the norm $\left\|\phi_{NL}^{k+1}-\phi_{NL}^{k} \right\|_{\infty}$ with nonlinear iteration count ($I_{NL}$) for different schemes using Picard and Newton-Raphson methods on 1AJF.}
    \label{fig:Norm_General_vs_Iter}
\end{figure}

Tables \ref{table:Picard_Vec_C} and \ref{table:Newton-Raphson_Vec_C} confirm that both the Picard and Newton-Raphson converge robustly to the identical value for the nonlinear solvation energy contribution of the 1AJF system. The Newton-Raphson method demonstrates superior convergence characteristics when using a constant relaxation factor ($\omega^{con}$), requiring an iteration count less than half that of Picard, which results in an overall reduction in total runtime of approximately 40\%. This accelerated convergence is visually confirmed in Figure~\ref{fig:Norm_General_vs_Iter} by the significantly faster decay of the residual norm. 

Analysis of the computational overhead provides context for this speed-up: the Newton-Raphson iteration achieves a marked 60\% reduction in the Nonlinear Solver Time (NIS). While this gain is partially mitigated by a 10\% increase in the Preprocessing Time (PT) due to the calculation of the Jacobian. Since the accelerated NIS component now accounts for only one-third of the total solving time, the dominance of this efficiency gain successfully drives the overall ~40\% performance improvement.

\begin{figure}
    \centering
    \includegraphics[width=0.7\textwidth]{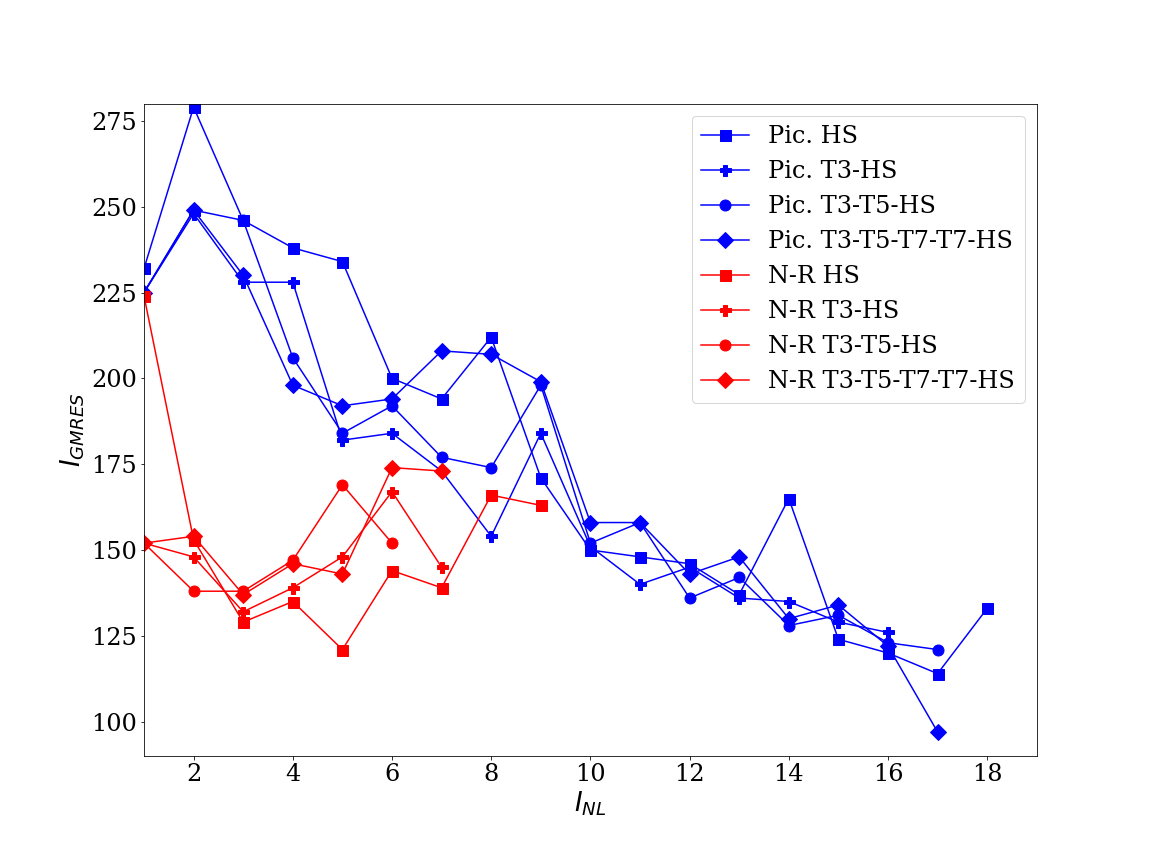}
    \caption{Number GMRES iterations in each nonlinear iteration ($I_{NL}$) for different schemes using Picard and Newton-Raphson methods on 1AJF.}
    \label{fig:Gmres_General_vs_Iter}
\end{figure}

Figure \ref{fig:Gmres_General_vs_Iter} compares the GMRES iteration count required at each nonlinear step. For Picard iterations, the linear solution from the previous step serves as an effective start for the subsequent linear solver. Consequently, as the nonlinear residual decays, the number of required GMRES iterations steadily decreases. In contrast, the linear systems in subsequent Newton-Raphson steps are functionally independent, dictated by the continually changing Jacobian. Therefore, the GMRES iteration count does not decrease but remains stable, typically requiring around 150 iterations per nonlinear step throughout the convergence process.

Table \ref{table:Picard_Vec_C} shows that for Picard, the T3–HS scheme yields the shortest runtimes, closely followed by T3-T5-HS, which required an additional nonlinear iteration. Similarly, for Newton-Raphson in Table \ref{table:Newton-Raphson_Vec_C}, T3-T5-HS is slightly faster (by one iteration). To use the same approaches for both methods, we will continue our analysis with the simpler T3-HS scheme.  


Although we found an effective $\omega^{con}$ that minimises the the runtime for both iterative approaches, Tables \ref{table:Picard_Mul_omega} and \ref{table:Newton-Raphson_Mul_omega} show the iteration count is very sensitive to the choice of $\omega^{con}$, making desirable to automatically identify the optimal $\omega_k^{opt}$. In practice, manually selecting this value can be computationally expensive (due to the time required for testing) and may lead to very slow convergence.



\begin{table}[ht] 
\centering
\footnotesize
\caption{1AJF calculations using Picard nonlinear iterations with different constant relaxation factors $\omega^{con}$, under the T3–HS scheme.}
\begin{tabular}{|c|c|c|c|c|c|c|c|c|} 
\hline
 & \multicolumn{4}{c|}{\textbf{ESE [kcal/mol]}} & \multicolumn{3}{c|}{\textbf{Time [s]}} & \\
\cline{2-8}
\textbf\textbf{$\omega^{con}$} & \textbf{Total} & \textbf{LC} & \textbf{NC} & \textbf{$\Delta G_{\Omega_{i}}$}  & \textbf{Total}  & \textbf{PT} & \textbf{NIS} & \textbf{I}  \\
\hline
0.13  & -1128.994 & -1122.647 & -9.470 & 3.122 & 196.61	& 62.36	& 129.63 & 22 \\
0.15 & -1128.996 & -1122.647 & -9.471 & 3.122 & 177.20	& 57.45	& 115.41 & 20 \\
0.17 & -1128.997 & -1122.647 & -9.472 & 3.122 & 166.00	& 58.03	& 103.57 & 18 \\
0.21 & -1128.999 & -1122.647 & -9.474 & 3.121 & 162.74	& 60.83	& 97.15 & 16 \\
0.23 & -1129.003 & -1122.647 & -9.477 & 3.121 & 171.21	& 59.51	& 107.16 & 21 \\
0.25 & -1129.006 & -1122.647 & -9.479 & 3.120 & 407.13	& 62.19	& 340.25 & 64 \\
\hline
\end{tabular}
\label{table:Picard_Mul_omega}
\end{table}


\begin{table}[ht] 
\centering
\footnotesize
\caption{1AJF calculations using Newton-Raphson nonlinear iterations with different constant values of $\omega^{con}$, under the T3–HS scheme.}
\begin{tabular}{|c|c|c|c|c|c|c|c|c|} 
\hline
 & \multicolumn{4}{c|}{\textbf{ESE [kcal/mol]}} & \multicolumn{3}{c|}{\textbf{Time [s]}} & \\
\cline{2-8}
\textbf{$\omega$} & \textbf{Total} & \textbf{LC} & \textbf{NC} & \textbf{$\Delta G_{\Omega_{i}}$}  & \textbf{Total}  & \textbf{PT} & \textbf{NIS} & \textbf{I}  \\
\hline
1.7	& -1129.004	& -1122.647	& -9.478 & 3.121 & 189.66 & 64.56 & 121.30 & 20 \\
1.5	& -1129.004	& -1122.647	& -9.478 & 3.121 & 148.81 & 66.96 & 77.90 & 11 \\
1.3	& -1129.004	& -1122.647	& -9.478 & 3.121 & 115.76 & 64.35 & 47.77 & 8 \\
1.1	& -1129.004	& -1122.647	& -9.478 & 3.121 & 109.21 & 64.14 & 41.12 & 7 \\
1.0 & -1129.004 & -1122.647 & -9.478 & 3.121 & 113.37 & 65.99 & 43.75 &	7 \\
0.9	& -1129.004	& -1122.647	& -9.478 & 3.121 & 127.20 & 66.86 & 56.60 & 9 \\
0.7	& -1129.004	& -1122.647	& -9.478 & 3.121 & 144.37 & 64.83 & 76.02 & 13 \\
0.5	& -1129.004	& -1122.647	& -9.478 & 3.121 & 171.97 & 66.26 & 102.22 & 20 \\
\hline
\end{tabular}
\label{table:Newton-Raphson_Mul_omega}
\end{table}




\subsubsection{Comparison of  numerical methods to obtain $\omega_k^{opt}$.}


In this section, we perform calculations on 1AJF using the T3-HS scheme for $\mathbf{N}$, $Tol_{GR}=10^{-6}$, $Tol_{NL}=10^{-3}$, and a variable relaxation factor throughout iterations. At each iteration, we solve Equation~\eqref{eq:fd=0} to determine $\omega_k^{opt}$ to a tolerance of $Tol_{\omega}=10^{-2}$ using: Newton-Raphson (NR), Bisection (B), Bisection in the first iteration and Secant method in the rest (B-SEC), and B-SEC approximating the secant method with Equation~\eqref{eq:SEC_FD} when the residual is small (B-SEC(DF)). B-SEC(DF) is only applicable to Picard. More details can be found in~\ref{sec:one_dim_root_finding}. 


%
Except for the Bisection method, all other algorithms require initial guesses to start with iterations. For Picard with B-SEC and B-SEC(DF), we used $\omega_{k}^{-1}=1$ and $\omega_{k}^{0}=1.125$, and searched for $\omega_k^{opt}$ between 0 and 1 if the potential was calculated with Picard and between 0 and 4 when Newton-Raphson was applied. The Picard method with Newton-Raphson for $\omega_k^{opt}$ used 0.5 as initial guess in the first non-linear iteration (approximated with T3), and the previous $\omega_k^{opt}$ in subsequent iterations. 

By experience, we saw that $\omega_k^{opt}$ for the first Newton-Raphson iteration for the potential (approximated with T3) was consistently close to 2, regardless of the analysed system, whereas the rest of the iterations (with the full hyperbolic sine, HS) showed $\omega_k^{opt}$ closer to 1. This is why we chose those values as initial guesses of the Newton-Raphson scheme for $\omega_k^{opt}$ when Newton-Raphson was used to compute the potential.

\begin{table}[ht]
\centering
\footnotesize
\caption{1AJF calculations with Picard for the nonlinear iterations and a variable $\omega$, obtained with different methods to a tolerance of $Tol_{\omega}=10^{-2}$.}. 
\begin{tabular}{|c|c|c|c|c|c|c|c|c|} 
\hline
\textbf{Method} & \multicolumn{4}{c|}{\textbf{ESE [kcal/mol]}} & \multicolumn{3}{c|}{\textbf{Time [s]}} & \\
\cline{2-8}
\textbf{for $\omega$} & \textbf{Total} & \textbf{LC} & \textbf{NC} & \textbf{$\Delta G_{\Omega_{i}}$}  & \textbf{Total}  & \textbf{PT} & \textbf{NIS} & \textbf{I}  \\
\hline
B &	-1129.004 & -1122.647 & -9.478 & 3.121 & 396.54	& 60.28	& 331.93 & 13 \\
NR & -1129.003 & -1122.647 & -9.478 & 3.121 & 253.40 & 58.50 & 190.90 & 12 \\
B-SEC & -1129.003 & -1122.647 & -9.478 & 3.121 & 221.13	& 57.35	& 159.56 & 12 \\
B-SEC(DF) & -1129.003 & -1122.647 & -9.478 & 3.121 & 206.34	& 58.48	& 143.55 & 12 \\
\hline
\end{tabular}
\label{table:Picard_w_variable}
\end{table}


\begin{table}[ht]
\centering
\footnotesize
\caption{1AJF calculations with Newton-Raphson for the nonlinear iterations and a variable $\omega$, obtained with different methods to a tolerance of $Tol_{\omega}=10^{-2}$.}. 
\begin{tabular}{|c|c|c|c|c|c|c|c|c|} 
\hline
\textbf{Method} & \multicolumn{4}{c|}{\textbf{ESE [kcal/mol]}} & \multicolumn{3}{c|}{\textbf{Time [s]}} & \\
\cline{2-8}
\textbf{for $\omega$} & \textbf{Total} & \textbf{LC} & \textbf{NC} & \textbf{$\Delta G_{\Omega_{i}}$}  & \textbf{Total}  & \textbf{PT} & \textbf{NIS} & \textbf{I}  \\
\hline
B & -1129.004 & -1122.647 & -9.478 & 3.121 & 177.07	& 65.21	& 108.44 & 4 \\
NR & -1129.004 & -1122.647 & -9.478 & 3.121 & 110.26 & 63.40	& 43.79 & 4 \\
B-SEC & -1129.004 & -1122.647 &-9.478 & 3.121 & 136.76 & 65.82 & 67.44 & 4 \\
\hline
\end{tabular}
\label{table:Newton-Raphson_w_variable}
\end{table}



Tables \ref{table:Picard_w_variable} and \ref{table:Newton-Raphson_w_variable} present the comparison of the above-mentioned methods for Picard and Newton-Raphson, respectively. These tables show that the method is effective in finding an $\omega_k^{opt}$ in each iteration, reducing the best constant $\omega^{con}$ case from 16 to 12 iterations for Picard (Table \ref{table:Picard_Mul_omega} versus Table \ref{table:Picard_w_variable}) and from 7 to 4 iterations for Newton-Raphson (Table \ref{table:Newton-Raphson_Mul_omega} versus Table \ref{table:Newton-Raphson_w_variable}). However, timing comparisons show that once an appropriate constant $\omega^{con}$ is found (Tables \ref{table:Picard_Mul_omega} and \ref{table:Newton-Raphson_Mul_omega}), it is faster than to find $\omega_k^{opt}$ every iteration because of the time required to automatically calculate it. 

According to Tables \ref{table:Picard_w_variable} and \ref{table:Newton-Raphson_w_variable} the fastest algorithm to solve Equation \eqref{eq:fd=0} for $\omega_k^{opt}$ is the hybrid bisection-secant with direct formulation method (B-SEC(DF)) if Picard is used for the nonlinear iterations. In contrast, when the Newton–Raphson scheme is employed for the nonlinear iterations of the PBE, the optimal performance is achieved by also using Newton–Raphson to determine $\omega_k^{opt}$. It is worth noting that the computational time obtained with this fully Newton–Raphson strategy is only marginally higher than that observed with $\omega^{con}$. 

The shape of $f_d(\omega)$ in Equation \eqref{eq:fd=0} is key to understand the behaviour of the different methods. In the case of Picard, Figure \ref{fig:fd_esfera} shows a linear $f_d(\omega)$ and the starting point is always (except for the first iteration) set at $f_d(0)=1$, which maps perfectly to the Secant method. Then, the B–SEC method applies Bisection only for the first nonlinear iteration to handle the transition in the argument of $\Vcc$, which requires fewer internal iterations $I_{\omega}$ than Newton–Raphson at this stage. This happens because when the $f_d$ curve is not sufficiently smooth, Newton–Raphson demands significantly more iterations than Bisection. As for the Secant method, when initialized with $\omega_{k}^{-1}=0$ and $\omega_{k}^{0}=0.125$ in Equation~\eqref{eq:Met_Secante}, it computes $\omega^{opt}_k$ faster and in less iterations than Newton–Raphson. Moreover, computation times are further improved when the Secant method is used with the direct formulation given in Equation~\eqref{eq:SEC_FD}, especially for small tolerances ($Tol_{\omega}=10^{-2}$). Thus, this final approach proves to be the most efficient strategy for computing the optimal $\omega_k^{opt}$ within the nonlinear Picard method.

For Newton-Raphson, Figure \ref{fig:fd_esfera} also shows a linear $f_d(\omega)$ in all iterations except the first one; however, this time $f_d(0)$ varies across iterations. Then, the Newton-Raphson method becomes a good alternative to find $\omega_k^{opt}$. 

\subsection{Extension to highly charged biomolecules.} 

Having found optimal algorithmic choices with 1AJF, we now extend our calculations to other biomolecules with high charge using Newton-Raphson as the nonlinear solver with the T3-HS scheme for $\mathbf{N}$, and computing the $\omega_{k}^{opt}$ with Newton-Raphson. In addition to 1AJF, we considered the RNA-based molecules with PDB code 1RNA \cite{1rna}, 1TRA \cite{1tra}, 3WBM \cite{3wbm}, and 1HC8 \cite{1ch8}, which contain 879, 1997, 7227, and 5879 atoms, and a total charge of -25.4, -61, -16.6, -106 $[e]$, respectively. We downloaded the corresponding PDB files from the Protein Data Bank, and parametrised them with the \texttt{AMBER} force field \cite{ponder2003force} using the \texttt{PDB2PQR} software \cite{dolinsky2004pdb2pqr}.

\begin{table}[ht]
\centering
\footnotesize
\caption{Vertex and element information for the surface and volumetric meshes of 5 molecules. Note that this is a different mesh refinement than the one used on 1AJF to find the optimal algorithm.}
\begin{tabular}{|c|c|c|c|c|c|} 
\hline
\textbf{Molecule} & \multicolumn{4}{c|}{\textbf{Surface Mesh}} & \textbf{Vol.}\\
\cline{2-5}
 & \multicolumn{2}{c|}{$\Gamma_m$} & \multicolumn{2}{c|}{$\Gamma_s$} &  \textbf{Mesh}\\
\cline{2-5}
  & \textbf{Vert.} &  \textbf{Elem.} & \textbf{Vert.} & \textbf{Elem.} & \textbf{Vert.} \\ 
\hline
1AJF & 10218 & 20432 & 3674	& 7344 & 37303 \\
1RNA & 15874 & 31744 & 4856	& 9708 & 56940 \\
1TRA & 36624 & 73256 & 10760 & 21508 & 130891 \\
3WBM & 82670 & 165392 & 18286 & 36568 & 287897 \\
1HC8 & 71986 & 143984 & 16864 & 33724 & 252246 \\
\hline
\end{tabular}
\label{table:Informacion_5Moleculas}
\end{table}

The meshes follow the guidelines of the refinement study in \ref{sec:mesh}. We generated the volumetric mesh with \texttt{Tetmesh} (growth factor of 1.1) starting from the corresponding surface meshes using \texttt{Nanoshaper} with 4 vertices per \AA$^{2}$ and 1 vertices per \AA$^{2}$ for $\Gamma_m$ and $\Gamma_s$, respectively, separated 3 \AA. Table \ref{table:Informacion_5Moleculas} summarises the number of vertices and elements for both surface and volumetric meshes associated with each molecule. For the construction of the boundary operators, dense assembly was used for 1AJF and 1RNA, while the remaining cases were handled using the FMM.



\begin{table}[ht]
\centering
\footnotesize
\caption{Calculations on highly charged molecules.}
\begin{tabular}{|c|c|c|c|c|c|c|c|c|} 
\hline
 & \multicolumn{4}{c|}{\textbf{ESE [kcal/mol]}} & \multicolumn{3}{c|}{\textbf{Time [s]}} &  \\
\cline{2-8}
\textbf{Molecule} & \textbf{Total} & \textbf{LC} & \textbf{NC} & \textbf{$\Delta G_{\Omega_{i}}$}  & \textbf{Total}  & \textbf{PT} & \textbf{NIS} & \textbf{I}  \\
\hline
1AJF  & -1116.479 & -1110.090 & -9.479 & 3.090 & 164.00 & 	87.78 & 70.09 & 4 \\
1RNA  & -2150.037 & -2135.832 & -20.105	& 5.901	& 291.68 & 	154.87 & 125.17	& 5 \\
1TRA & -7804.765 & -7765.564 & -54.815 & 15.614	& 1551.68 & 396.06 & 1106.66 & 5 \\
3WBM  & -3079.227 & -3060.434 & -28.705	& 9.912	& 3685.60 & 1016.28	& 2519.19 & 5 \\
1HC8 & -21034.259 & -20813.168 & -268.016 & 46.924 & 3777.87 & 841.43 & 2832.03 & 6 \\
\hline
\end{tabular}
\label{table:5Moleculas}
\end{table}

The calculations that led to Table \ref{table:5Moleculas} used $Tol_{GR} = 10^{-6}$, $Tol_{NL}=10^{-3}$, and $Tol_\omega = 10^{-2}$. These results show that the proposed nonlinear scheme is robust with respect to size and charge. The same algorithmic choices converged calculations from 1AJF to 1HC8, which span in size and charge by 4$\times$, increasing only from 4 to 6 Newton-Raphson iterations. Despite being smaller than 3WBM, the highly charged 1HC8 system requires more nonlinear iterations to reach convergence. This increase reflects the stronger nonlinearity of the potential field. However, the total runtime remains close to that of 3WBM, as the smaller system size reduces the preprocessing and matrix assembly times, partially compensating for the additional iterations.

\ref{sec:small_optimisations} details other optimisation strategies that for the 1AJF model problem provide improvements in timing. These are:
\begin{enumerate}
    \item Adjustable GMRES tolerance as nonlinear iterations evolve, starting with an initial tolerance $Tol_{GR_0}$=10$^{-3}$.
    \item Skip the computation of $\omega^{opt}_k$ when the residual of the nonlinear solver is smaller than 0.35, and reuse the last calculated $\omega^{opt}_k$ for the subsequent iterations.
    \item Reduce the accuracy of the tolerance of $\omega^{opt}_k$ to $Tol_{\omega}=0.1$.
\end{enumerate}
Table \ref{table:5Moleculas_optimised} presents calculations of the solvation free energy for the same highly-charged molecules as Table \ref{table:Informacion_5Moleculas}, but incorporating the optimisations from~\ref{sec:small_optimisations}. Although for 1AJF these optimisations yield a speed up of only 1.12$\times$ (from $103.64 s$ for $\omega^{con}$ to $91.95 s$ when including all optimisations), they become more important as the charge increases, giving a performance speed up of 1.37$\times$ for 1HC8. 

\begin{table}[ht]
\centering
\footnotesize
\caption{Calculations on highly charged molecules with optimisations detailed in \ref{sec:small_optimisations}.}
\begin{tabular}{|c|c|c|c|c|c|c|c|c|} 
\hline
 & \multicolumn{4}{c|}{\textbf{ESE [kcal/mol]}} & \multicolumn{3}{c|}{\textbf{Time [s]}} &  \\
\cline{2-8}
\textbf{Molecule} & \textbf{Total} & \textbf{LC} & \textbf{NC} & \textbf{$\Delta G_{\Omega_{i}}$}  & \textbf{Total}  & \textbf{PT} & \textbf{NIS} & \textbf{I}  \\
\hline
1AJF  & -1116.479 & -1110.215 & -9.354 & 3.090 & 146.52 & 82.78 & 57.49 & 5 \\
1RNA  & -2150.037 & -2135.688 & -20.249	& 5.901	& 236.23 &  140.69 & 84.77 & 5 \\
1TRA & -7804.765 & -7764.870 & -55.510 & 15.614	& 1086.18 & 261.92 & 776.76	& 5 \\
3WBM  & -3079.227 & -3061.357 & -27.782	& 9.912	& 2609.27 & 716.84 & 1744.23 & 5 \\
1HC8  & -21034.259 & -20790.436	& -290.748 & 46.924	& 2737.08 & 578.80 & 2055.31 & 6 \\
\hline
\end{tabular}
\label{table:5Moleculas_optimised}
\end{table}

As mentioned earlier, we used $\omega_0^0=2$ as the initial guess to find $\omega_0^{opt}$ (in the first nonlinear iteration). This results in a large overhead for 1HC8, which takes 10 Newton-Raphson iterations to find $\omega_0^{opt}$ to $Tol_\omega=0.1$, motivating a final optimisation strategy where the Bisection and Secant method are used to obtain a better initial guess. 

We start by obtaining three values of $\omega$ with Bisection ($\omega_{in}^{0}$, $\omega_{in}^{1}$, and $\omega_{in}^{2}$) in the range between 2 and 3, along the corresponding evaluations $f_d(\omega_{in}^{0})$, $f_d(\omega_{in}^{1})$, and $f_d(\omega_{in}^{2})$. Starting from these three values, we iteratively use the Secant method adapting Equation \eqref{eq:Met_Secante} as:
\begin{equation}
    \label{eq:Met_Secante}
    \omega_{in}^{j+1}=\omega_{in}^{j-1}-\frac{(\omega_{k}^{j-1}-\omega_{in}^{j})f_d(\omega_{in}^{j-1})}{f_d(\omega_{in}^{j-1})-f_d(\omega_{in}^{j})}
\end{equation}
until $\left| \omega_{in}^{j+1}-\omega_{in}^{j}\right|<0.05$ to set the initial guess $\omega_0^0 = \omega_{in}^{j+1}$. Table \ref{table:1hc8_improved} shows that this optimisation provides an extra 5\% increase in performance for 1HC8, however, it may become more important for systems with higher charge.

\begin{table}[ht]
\centering
\footnotesize
\caption{Calculations on 1HC8 with optimisations detailed in \ref{sec:small_optimisations} and using Bisection and Secant methods to find the initial guess of $\omega$ in the first nonlinear iteration.}
\begin{tabular}{|c|c|c|c|c|c|c|c|c|} 
\hline
 & \multicolumn{4}{c|}{\textbf{ESE [kcal/mol]}} & \multicolumn{3}{c|}{\textbf{Time [s]}} &  \\
\cline{2-8}
\textbf{Molecule} & \textbf{Total} & \textbf{LC} & \textbf{NC} & \textbf{$\Delta G_{\Omega_{i}}$}  & \textbf{Total}  & \textbf{PT} & \textbf{NIS} & \textbf{I}  \\
\hline
1HC8 & -21034.259 & -20790.436 & -290.748 & 46.924 & 2596.32	& 571.31 & 1918.79 & 6 \\
\hline
\end{tabular}
\label{table:1hc8_improved}
\end{table}

%% file: Conclusiones_Principal.tex
\section{Conclusion}

In this work, we proposed a methodology to solve the nonlinear PBE with a coupled FEM-BEM scheme, which models electrostatics in highly charged molecular solvation. The main step is to separate the total potential into linear and nonlinear contributions, where the latter is solved with Picard or Newton-Raphson iterations that gradually incorporate the nonlinearity with a Taylor expansion. We saw that Newton-Raphson was the most efficient approach to solve the PBE, which agrees with previous works~\cite{holst1995numerical,cai2010performance}. Solvation energy results for a sphere are consistent with APBS calculations, validating our approach. Then, we used 1AJF, an RNA structure, to perform a detailed analysis of the different algorithmic choices ensure convergence and efficiency.

Our technique uses a cubic Taylor expansion to approximate the nonlinearity in the first iteration, followed by the full hyperbolic sine in subsequent steps (T3-HS scheme). While this approach is effective, its efficiency depends heavily on the manually-selected relaxation parameter, which can limit usability. We address this issue by introducing a method that automatically determines the optimal relaxation parameter at each iteration, both reducing the number of nonlinear iterations and ensuring convergence. The optimal parameter is found by solving a one-dimensional nonlinear equation via Newton-Raphson. As a result, our basic implementation is as fast as the best manually tuned method, but does not require manual parameter selection.


We also present a set of optimisations that further decrease the time-to-solution by: $(i)$ limiting the number of times the optimal relaxation factor is calculated, $(ii)$ using a high tolerance of the nonlinear solver for the relaxation parameter, and $(iii)$ adjusting the GMRES solver tolerance as the nonlinear iterations progress. These optimisations result in a 1.37$\times$ speedup for the molecule with the highest charge. 

Our results further demonstrate that even modest improvements in the initial iterations can significantly enhance the efficiency of Newton-based methods. This effect is particularly pronounced when Newton iterations are susceptible to initial stagnation. In these situations, solving the one-dimensional problem with a different method, can yield a substantially important reduction in overall solver time.

In future work, we aim to extend this methodology to solve other types of nonlinear equations. Additionally, we plan to apply our approach to investigate the role of the nonlinearity in molecular solvation, for example, in the surface potential, which plays a crucial role in determining binding affinities in drug design.

%% file: Appendices.tex
\section{Methods to find an one dimensional root}
\label{sec:one_dim_root_finding}

This appendix describes the implementation of Bisection, Newton–Raphson, and the Secant method to find the optimal relaxation factor for iteration $k$ ($\omega^{opt}_{k}$).  The superscript $j$ denotes the iteration index used to approximate \(\omega^j_{k}\) until $f_d(\omega_{k}^{j})<Tol_{\omega}$, and we assign $\omega^{opt}_{k}=\omega_{k}^{j}$.

The Bisection method 
repeatedly halves an initial search interval and selects the subinterval where a sign change occurs. This method is particularly useful when the function exhibits large gradients.

When the function is sufficiently smooth, the Newton–Raphson method offers faster convergence with fewer iterations compared to Bisection. However, its convergence strongly depends on selecting a suitable initial guess $\omega_{k}^{0}$. 
The Newton–Raphson method updates for $\omega^j_{k}$ is given by:
\begin{equation*}
    \omega_{k}^{j+1}=\omega_{k}^{j}-\frac{f_d(\omega_{k}^{j})}{\partial_{\omega_{k}^{j}}f_d(\omega_{k}^{j})}.
\end{equation*}
where, the derivative $\partial_{\omega_{k}^{j}}f_d(\omega_{k}^{j})$ is calculated as follows:
\begin{align*}
      \partial_{\omega_{k}}f_d(\omega_{k})&=\partial F_{1}+\partial F_{2} -1 \\ \nonumber
        \partial F_{1}&=-\frac{(\za+\zdb)\cdot\zdb+\zb\cdot\zddb}{(\za+\zdb )\cdot\za} \\   \nonumber
        \partial F_{2}&= \frac{((\za+\zdb)\cdot\zb)(\za\cdot\zddb)}{((\za+\zdb  )\cdot\za)^{2}} 
\end{align*}   
where the auxiliary terms are defined
    in Equation \eqref{eq:ab_pic} and:
\begin{equation*} 
    \zddb  = \left[\zc\right]^{2}\partial_{\Phi}^2\Vc 
\end{equation*}

The Secant method is a variation of Newton–Raphson method in which the derivative $f_d(\omega_{k}^{j})$ is not computed explicitly. Instead, it uses the finite difference approximation, 
and the update rule for \(\omega^j_{k}\) is:
\begin{equation}
    \label{eq:Met_Secante}
    \omega_{k}^{j+1}=\omega_{k}^{j-1}-\frac{(\omega_{k}^{j-1}-\omega_{k}^{j})f_d(\omega_{k}^{j-1})}{f_d(\omega_{k}^{j-1})-f_d(\omega_{k}^{j})}
\end{equation} 

Figures~\ref{fig:gd_esfera} and~\ref{fig:fd_esfera} show that for Picard iterations, \(f_d(0)=1\) in all iterations, except the first one. Considering $\mathbf{N}$ remains unchanged throughout iterations, we can use this fact to our advantage, and introduce the initial value 
$\omega_{k}^{-1}=0$ and $f_d(\omega_{k}^{-1})=1$. 


Finally, when \(f_d\) behaves approximately linearly and the root is already found in the first iteration (i.e., $f_d(\omega_{k}^{1})<Tol_{\omega}$ using $\omega_{k}^{-1}=0$), the secant method simplifies further into a direct formulation:
\begin{equation}
    \label{eq:SEC_FD}
    \omega^{opt}_{k}=\frac{\omega_{k}^{0}}{1-f_d(\omega_{k}^{0})}
\end{equation}

\section{Other optimisation strategies.\label{sec:small_optimisations}}

This appendix describes three optimisation strategies that provided a minor improvement in performance for the 1AJF model problem. However, we cover them because they may become important for systems with larger size or charge. The analysis is done only for nonlinear iterations using 
\begin{itemize}
    \item Picard method with the T3-HS scheme, and B-SEC(DF) for the solver to find the $\omega_k^{opt}$.
    \item Newton-Raphson with the T3-HS scheme, and Newton-Raphson for the solver to find the $\omega_k^{opt}$.
\end{itemize}

\subsection{Reducing number of times $\omega$ is calculated}

The numerical computation of $\omega^{opt}_k$ is an added cost compared to using a constant $\omega^{con}$.  Therefore, an effective strategy to optimise computational time is to minimise the number of times $\omega^{opt}_k$ is computed, and reuse it as much as possible. In Picard, we computed $\omega^{opt}_k$ depending on the norm ratio $\frac{e_{k-1}}{e_k}$ of two subsequent iterations. If the ratio falls below a threshold $R_e$, it indicates that the nonlinear solver is diverging or converging too slowly, and a new $\omega^{opt}_k$ must be recomputed to restore efficiency. 

\begin{table}
\centering
\footnotesize
\caption{1AJF calculations with Picard for different norm ratio values $R_e$.}
\begin{tabular}{|c|c|c|c|c|c|c|c|c|} 
\hline
 & \multicolumn{4}{c|}{\textbf{ESE [kcal/mol]}} & \multicolumn{3}{c|}{\textbf{Time [s]}} & \\
\cline{2-8}
\textbf{$R_e$} & \textbf{Total} & \textbf{LC} & \textbf{NC} & \textbf{$\Delta G_{\Omega_{i}}$}  & \textbf{Total}  & \textbf{PT} & \textbf{NIS} & \textbf{I}  \\
\hline
1.3 & -1128.998	& -1122.647	& -9.473 & 3.122 & 181.80 & 58.36 & 119.12 & 18 \\
1.4 & -1129.002	& -1122.647	& -9.476 & 3.121 & 153.07 & 57.08 & 91.80 & 11 \\
1.5 & -1129.002	& -1122.647	& -9.476 & 3.121 & 160.38 & 62.62 & 93.37 & 11 \\
1.8 & -1129.003 & -1122.647	& -9.477 & 3.121 & 173.42 & 62.52 & 107.07 & 13 \\
\hline
\end{tabular}
\label{table:Picard_w_variable_con_Re}
\end{table}

Table \ref{table:Picard_w_variable_con_Re} shows that $R_e$ is an effective way to decrease the number of calculations of $\omega^{opt}_k$ without sacrificing the total count of non-linear iterations. The sweet spot is $R_e$ = 1.4, which decreases the time-to-solution making it faster than using a constant hand-picked $\omega^{con}$ (Table \ref{table:Picard_Mul_omega}), improving performance by $\sim$5\%.

In the case of the Newton-Raphson method, we propose to skip the calculation of $\omega_k^{opt}$ when the residual of the nonlinear solver falls below a threshold $L_e$. For all subsequent iterations, the last optimal $\omega_k^{opt}$ is reused.

Table \ref{table:NR_w_variable_con_Le} presents timings for different choices of $L_e$ using the best-performing calculation in Table \ref{table:Newton-Raphson_w_variable} as base. Note from Table \ref{table:NR_w_variable_con_Le} that if $L_e$ is too small, the algorithm searches for the optimal $\omega$ in every iteration, producing the same timing as Table \ref{table:Newton-Raphson_w_variable}. On the other hand, a high $L_e$ considerably increases the iteration count. For 1AJF we found that $L_e=0.35$ reduces the number of calculations for $\omega$ without affecting the total nonlinear iteration count, improving performance by $\sim$4\%.
\begin{table}
\centering
\footnotesize
\caption{1AJF calculations with Newton-Raphson for different values of $L_e$.}
\begin{tabular}{|c|c|c|c|c|c|c|c|c|} 
\hline
 & \multicolumn{4}{c|}{\textbf{ESE [kcal/mol]}} & \multicolumn{3}{c|}{\textbf{Time [s]}} & \\
\cline{2-8}
\textbf{$L_e$} & \textbf{Total} & \textbf{LC} & \textbf{NC} & \textbf{$\Delta G_{\Omega_{i}}$}  & \textbf{Total}  & \textbf{PT} & \textbf{NIS} & \textbf{I}  \\
\hline
0.04 & -1129.004 & -1122.647 & -9.478 & 3.121 & 110.26 & 63.40 & 43.79 & 4 \\
0.10 & -1129.004 & -1122.647 & -9.478 & 3.121 & 106.61 & 62.94 & 40.27 & 4 \\
0.35 & -1129.004 & -1122.647 & -9.478 & 3.121 & 105.46 & 61.30 & 40.83 & 4 \\
0.50 & -1129.004 & -1122.647 & -9.478 & 3.121 & 114.66 & 64.97 & 46.26 & 5 \\
6.00 & -1128.947 & -1122.647 & -9.446 & 3.146 & 196.47 & 63.39 & 129.52 & 20 \\
\hline
\end{tabular}
\label{table:NR_w_variable_con_Le}
\end{table}




\subsection{Choice of the tolerance $Tol_\omega$.}


High-precision is not required for $\omega$, as long as it is close enough to the optimal value.
Tables \ref{table:Picard_w_tolw_variable} and \ref{table:NR_w_tolw_variable} show energy calculations for Picard and Newton-Raphson, respectively, using the same setup as the best-performing cases in Tables \ref{table:Picard_w_variable_con_Re} (Picard, $R_e=1.4$) and \ref{table:NR_w_variable_con_Le} (Newton-Raphson, $L_e=0.35$) for different values of $Tol_\omega$. The lowest total runtime corresponds to a tolerance of $Tol_{\omega}=0.1$ in both cases, which is the highest value that does not increase the nonlinear iteration count, improving performance by $\sim$6.6\% for Picard and $\sim$3\% for Newton-Raphson.


\begin{table}[ht]
\centering
\footnotesize
\caption{1AJF calculations for different values of the tolerance $Tol_{\omega}$ using Picard.}
\begin{tabular}{|c|c|c|c|c|c|c|c|c|} 
\hline
 & \multicolumn{4}{c|}{\textbf{ESE [kcal/mol]}} & \multicolumn{3}{c|}{\textbf{Time [s]}} & \\
\cline{2-8}
\textbf{$Tol_{\omega}$} & \textbf{Total} & \textbf{LC} & \textbf{NC} & \textbf{$\Delta G_{\Omega_{i}}$}  & \textbf{Total}  & \textbf{PT} & \textbf{NIS} & \textbf{I}  \\
\hline
0.01 & -1129.002 & -1122.647 & -9.476 & 3.121 & 153.07 & 57.08	& 91.80	& 11 \\
0.05 & -1129.002 & -1122.647 & -9.476 & 3.121 & 149.95 & 59.90	& 85.72	& 11 \\
0.1 & -1129.003 & -1122.647 & -9.476 & 3.121 & 142.91 & 60.20	& 78.46	& 11 \\
0.5 & -1129.005 & -1122.647 & -9.478 & 3.120 & 148.15 & 58.35	& 85.63	& 13 \\
\hline
\end{tabular}
\label{table:Picard_w_tolw_variable}
\end{table}

\begin{table}[ht]
\centering
\footnotesize
\caption{1AJF calculations for different values of the tolerance $Tol_{\omega}$ using Newton-Raphson.}
\begin{tabular}{|c|c|c|c|c|c|c|c|c|} 
\hline
 & \multicolumn{4}{c|}{\textbf{ESE [kcal/mol]}} & \multicolumn{3}{c|}{\textbf{Time [s]}} & \\
\cline{2-8}
\textbf{$Tol_{\omega}$} & \textbf{Total} & \textbf{LC} & \textbf{NC} & \textbf{$\Delta G_{\Omega_{i}}$}  & \textbf{Total}  & \textbf{PT} & \textbf{NIS} & \textbf{I}  \\
\hline
0.01 & -1129.004 & -1122.647 & -9.478 & 3.121 & 105.46 & 61.30 & 40.83 & 4 \\
0.05 & -1129.004 & -1122.647 & -9.478 & 3.121 & 104.77 & 64.42 & 36.99 & 4 \\
0.1	& -1129.004	& -1122.647	& -9.478 & 3.121 & 101.94 & 62.80 & 35.57 & 4 \\
0.5	& -1129.004	& -1122.647	& -9.478 & 3.121 & 106.31 & 64.71 & 38.19 & 5 \\
\hline
\end{tabular}
\label{table:NR_w_tolw_variable}
\end{table}

\subsection{Adjustable GMRES tolerance} 

There is no need to solve the linear system to a tight tolerance in the initial nonlinear iterations, allowing us to use a high value of the GMRES solver tolerance ($Tol_{GR}$), and adjust it as the nonlinear solver progresses.
We used an initial tolerance ($Tol_{GR_0}$) to solve for the linear contribution (LC) and in the early nonlinear iterations. We multiplied the tolerance by 0.1 every time the condition $\left \lceil \frac{e_k}{1250 \cdot Tol_{GR_0}} \right \rceil < 1$ is met, until it reached the final value $Tol_{GR}$.

Tables \ref{table:Picard_w_tolw_tolg_variable} and  \ref{table:NR_w_tolw_tolg_variable} shows the ESE of 1AJF for different values of $Tol_{GR_0}$ with Picard and Newton-Raphson, respectively. The base case is the setup with $Tol_\omega=0.1$ in Tables \ref{table:Picard_w_tolw_variable} (Picard) and \ref{table:NR_w_tolw_variable} (Newton-Raphson). In both cases, using $Tol_{GR_0} = 10^{-3}$ gave a performance enhancement of $\sim$10\%.



\begin{table}[ht]
\centering
\footnotesize
\caption{1AJF calculations for different values of the initial GMRES tolerance $Tol_{GR_0}$ using Picard.}
\begin{tabular}{|c|c|c|c|c|c|c|c|c|} 
\hline
 & \multicolumn{4}{c|}{\textbf{ESE [kcal/mol]}} & \multicolumn{3}{c|}{\textbf{Time [s]}} & \\
\cline{2-8}
\textbf{$Tol_{GR_0}$} & \textbf{Total} & \textbf{LC} & \textbf{NC} & \textbf{$\Delta G_{\Omega_{i}}$}  & \textbf{Total}  & \textbf{PT} & \textbf{NIS} & \textbf{I}  \\
\hline
$10^{-5}$ & -1129.003 & -1122.647 & -9.476 & 3.121 & 131.72	& 57.27	& 70.33	& 11 \\
$10^{-4}$ & -1129.003 & -1122.611 & -9.512 & 3.121 & 133.69	& 59.07	& 69.98	& 11 \\
$10^{-3}$ & -1129.003 & -1121.965 & -10.159	& 3.121	& 127.42 & 55.95 & 67.27 & 11 \\
$10^{-2}$ & -1129.002 & -1121.165 & -10.958	& 3.121	& 130.61 & 57.04 & 69.02 & 11 \\
\hline
\end{tabular}
\label{table:Picard_w_tolw_tolg_variable}
\end{table}

\begin{table}[ht]
\centering
\footnotesize
\caption{1AJF calculations for different values of the initial GMRES tolerance $Tol_{GR_0}$ using Newton-Raphson.}
\begin{tabular}{|c|c|c|c|c|c|c|c|c|} 
\hline
 & \multicolumn{4}{c|}{\textbf{ESE [kcal/mol]}} & \multicolumn{3}{c|}{\textbf{Time [s]}} & \\
\cline{2-8}
\textbf{$Tol_{GR_0}$} & \textbf{Total} & \textbf{LC} & \textbf{NC} & \textbf{$\Delta G_{\Omega_{i}}$}  & \textbf{Total}  & \textbf{PT} & \textbf{NIS} & \textbf{I}  \\
\hline
$10^{-5}$ & -1129.004 & -1122.647 & -9.478 & 3.121 & 97.25 & 59.89 & 34.08 & 4 \\
$10^{-4}$ &-1129.004 & -1122.611 & -9.514 & 3.121 & 95.93 & 59.76 & 32.77 & 4 \\
$10^{-3}$ &-1129.004 & -1121.965 & -10.160	& 3.121	& 91.95 & 57.15 & 31.71	& 4 \\
$10^{-2}$ &-1129.004 & -1121.165 & -10.960	& 3.121	& 91.56 & 57.34 & 31.09 & 4 \\
\hline
\end{tabular}
\label{table:NR_w_tolw_tolg_variable}
\end{table}

According to Table \ref{table:NR_w_tolw_tolg_variable}, the fastest alternative is to use Newton-Raphson incorporating all algorithmic optimisations detailed in this Appendix, which results in a total time of 91.56 s. This is $\sim$1.2$\times$ faster than not using them (Table \ref{table:Newton-Raphson_w_variable}) or using a hand-picked $\omega^{con}$ (Table \ref{table:Newton-Raphson_Mul_omega}).

\section{Mesh studies for calculations on 1AJF.\label{sec:mesh}}



This appendix studies mesh parameters to ensure accuracy of the numerical solution of the PBE. Here, simulations use Picard with the T3-HS scheme as the nonlinear solver, B-SEC(DF) with $Tol_\omega=0.1$ for an optimal $\omega$, and $R_e=1.4$. We performed a mesh refinement analysis according to Table \ref{table:Informacion_1AJF} and studied the influence of the surface mesh densities and distance between $\Gamma_m$ and $\Gamma_s$ in timing and accuracy. The volumetric mesh, generated with \texttt{Tetmesh}, adjusted to the surface meshes and limited the growth factor between tetrahedra to 1.1.  

The boundary operators were assembled generating the full BEM dense matrix, except for the finest mesh (grid 8), which employed the Fast Multipole Method (FMM). FMM results use an expansion order of 3 and leaf boxes with less than 100 quadrature nodes inside.

\begin{table}[ht]
\centering
\footnotesize
\caption{Vertex and element information for mesh refinement study with 1AJF.}
\begin{tabular}{|c|c|c|c|c|c|} 
\hline
 & \multicolumn{4}{c|}{\textbf{Surface Mesh}} & \textbf{Vol.}\\
\cline{2-5}
\textbf{Vertex} & \multicolumn{2}{c|}{\textbf{$\Gamma_m$}} & \multicolumn{2}{c|}{\textbf{$\Gamma_s$}} & \textbf{mesh} \\
\cline{2-5}
\textbf{dens.}  & \textbf{Vert.} &  \textbf{Elem.} & \textbf{Vert.} & \textbf{Elem.} & \textbf{Vert.} \\ 
\hline
1 & 2446 & 4888	& 3674 & 7344  & 11969 \\
2 & 4764 & 9524	& 6926 & 13848  & 25284 \\
4 & 10218 & 20432 & 14756 & 29508  & 59564 \\
8 & 21246 & 42488 & 30546 & 61088 & 131373 \\
\hline
\end{tabular}
\label{table:Informacion_1AJF}
\end{table}

\begin{table}[ht]
\centering
\footnotesize
\caption{Mesh refinement study on the 1AJF structure.} 
\begin{tabular}{|c|c|c|c|c|c|} 
\hline
\textbf{Vertex} & \multicolumn{4}{c|}{\textbf{ESE [kcal/mol]}} & \\
\cline{2-5}
\textbf{dens.} & \textbf{Total} & \textbf{LC} & \textbf{NC} & \textbf{$\Delta G_{\Omega_{i}}$}  & \textbf{I}  \\
\hline
1 & -1153.879 & -1146.760 & -10.266	& 3.148	&  11 \\
2 & -1129.003 & -1121.965 & -10.159	& 3.121	&  11 \\
4 & -1116.477 & -1110.056 & -9.533 & 3.111 &  15 \\
8 & -1110.996 & -1104.746 & -9.338 & 3.088	& 11 \\
\hline
\end{tabular}
\label{table:1AJF_Ref_Malla_General}
\end{table}

\begin{table}[ht]
\centering
\footnotesize
\caption{Richardson extrapolation and observed order of convergence (o.o.c.) of the solvation energy for 1AJF from Table \ref{table:1AJF_Ref_Malla_General}.}
\begin{tabular}{|c|c|c|} 
\hline
& \textbf{ESE}  & \\
\textbf{Case} & \textbf{[kcal/mol]} & \textbf{o.o.c}  \\
\hline
1-2-4 & -1103.776 & 0.99\\
2-4-8 & -1106.988 & 1.19\\
\hline
\end{tabular}
\label{table:Extrapolacion_1AJF}
\end{table}


As shown in Table \ref{table:1AJF_Ref_Malla_General}, refining the surface meshes the solvation
energy approaches the value extrapolated via Richardson in Table \ref{table:Extrapolacion_1AJF} with an observed order of convergence that is very close to the theoretical first order.



As observed in the previous analyses, refining the surface mesh improves the accuracy of the solvation energy calculation. However, computational times can be further optimised by recognizing that not all regions of the domain require the same level of refinement.
In particular, the nonlinear contributions to the electrostatic potential are concentrated near the $\Gamma_{m}$ interface, the location of the permittivity jump. Therefore, a finer mesh near $\Gamma_{m}$ should be sufficient to ensure accuracy, while a coarser mesh on the outer interface $\Gamma_{s}$ should be acceptable without significantly affecting the final result.

To test this, we performed a selective mesh refinement analysis: the internal surface mesh ($\Gamma_{m}$) is fixed with a density of 4 vertices per \AA$^{2}$, while the refinement of $\Gamma_{s}$ is varied. The mesh configurations are detailed in Table \ref{table:Informacion_1AJF_V2}, and the results  are summarised in Table \ref{table:1AJF_Ref_Malla_Sup_Ext}.

Table \ref{table:1AJF_Ref_Malla_Sup_Ext} demonstrates that employing a mesh for $\Gamma_s$ that is four times coarser than that of $\Gamma_m$ preserves the solvation energy to within five significant digits of accuracy.


\begin{table}[ht]
\centering
\footnotesize
\caption{Vertex and element information for 1AJF meshes with varying density of the outer surface mesh, keeping the inner mesh fixed at 4 vertices per \AA$^2$.}
\begin{tabular}{|c|c|c|c|c|c|} 
\hline
 & \multicolumn{4}{c|}{\textbf{Surface Mesh}} & \textbf{Vol.}\\
\cline{2-5} 
\textbf{Vertex} & \multicolumn{2}{c|}{$\Gamma_m$} & \multicolumn{2}{c|}{$\Gamma_s$} & \textbf{mesh}  \\
\cline{2-5} 
\textbf{dens.}  & \textbf{Vert.} &  \textbf{Elem.} & \textbf{Vert.} & \textbf{Elem.} & \textbf{vert.} \\ 
\hline
0.5 & 10218 & 20432 & 2044 & 4084 & 34825 \\
1 & 10218 & 20432 & 3674 & 7344 & 37303 \\
2 & 10218 & 20432 & 6926 & 13848 & 43629 \\
4 & 10218 & 20432 & 14756 & 29508  & 59564 \\
\hline
\end{tabular}
\label{table:Informacion_1AJF_V2}
\end{table}

\begin{table}[ht]
\centering
\footnotesize
\caption{Results of 1AJF calculations with varying density of the outer surface mesh, keeping the inner mesh fixed at 4 vertices per \AA$^2$.}
\begin{tabular}{|c|c|c|c|c|c|} 
\hline
\textbf{Vertex} & \multicolumn{4}{c|}{\textbf{ESE [kcal/mol]}}  & \\
\cline{2-5}
\textbf{density} & \textbf{Total} & \textbf{LC} & \textbf{NC} & \textbf{$\Delta G_{\Omega_{i}}$}  &  \textbf{I}  \\
\hline
0.5	& -1116.592	& -1110.408	& -9.249 & 3.065 &  11 \\
1 & -1116.477 & -1110.215 & -9.352 & 3.090 &  11 \\
2 & -1116.457 & -1110.150 & -9.395 & 3.087 &  13 \\
4 & -1116.477 & -1110.056 & -9.533 & 3.111 &  15 \\
\hline
\end{tabular}
\label{table:1AJF_Ref_Malla_Sup_Ext}
\end{table}


\subsection{Surface mesh distances for 1AJF}
This section analyses the impact of the distance between the inner and outer surface meshes on the solvation energy. This analysis is motivated by the fact that nonlinear effects are concentrated near the inner interface $\Gamma_{m}$, where the electrostatic potential is high. As the distance from this interface increases, the potential (and thus the nonlinear contribution) decreases. The inner and outer surface meshes were generated using Nanoshaper with a fixed density of 4 vertices per \AA$^{2}$, while the distance between the two surfaces was varied as detailed in Table \ref{table:Informacion_1AJF_V3} ($x$ in Figure~\ref{fig:Radio_de_Prueba}). The volumetric mesh was generated using \texttt{Tetmesh}, applying a growth factor of 1.1 between adjacent tetrahedra.

We used a dense matrix assembler on \texttt{Bempp-cl}, except in the case of a 6 \AA~ distance, where FMM was required due to the large mesh size.



\begin{table}[ht]
\centering
\footnotesize
\caption{Vertex and element information for 1AJF meshes at different distances between surface meshes ($x$ in figure~\ref{fig:Radio_de_Prueba}), using a mesh density of 4 vertices per \AA$^2$.}
\begin{tabular}{|c|c|c|c|c|c|} 
\hline
 & \multicolumn{4}{c|}{\textbf{Surface Mesh}} & \textbf{Vol.}\\
\cline{2-5}
\textbf{Dist.} & \multicolumn{2}{c|}{\textbf{$\Gamma_m$}} & \multicolumn{2}{c|}{\textbf{$\Gamma_s$}} & \textbf{ Mesh}\\
\cline{2-5}
\textbf{\AA~}  & \textbf{Vert} &  \textbf{Elem} & \textbf{Vert} & \textbf{Elem} & \textbf{Vert} \\ 
\hline
2 & 10218 & 20432 & 13448 & 26892 & 52749 \\
3 & 10218 & 20432 & 14756 & 29508 & 59564 \\
4 & 10218 & 20432 & 16298 & 32592 & 64711 \\
5 & 10218 & 20432 & 18074 & 36144 & 69738 \\
6 & 10218 & 20432 & 19908 & 39812 & 74736 \\
\hline
\end{tabular}
\label{table:Informacion_1AJF_V3}
\end{table}

\begin{table}[ht]
\centering
\footnotesize
\caption{FEM–BEM coupling with optimised Picard iteration for 1AJF for different distances between surface meshes, with an inverse-length of 4.}
\begin{tabular}{|c|c|c|c|c|c|} 
\hline
\textbf{Dist.} & \multicolumn{4}{c|}{\textbf{ESE [kcal/mol]}} & \\
\cline{2-5}
\textbf{\AA} & \textbf{Total} & \textbf{LC} & \textbf{NC} & \textbf{$\Delta G_{\Omega_{i}}$} & \textbf{I}  \\
\hline
2 & -1116.213 & -1110.138 & -9.221 & 3.147 & 12 \\
3 & -1116.477 & -1110.056 & -9.533 & 3.111 & 15 \\
4 & -1116.643 & -1110.219 & -9.499 & 3.075 & 12 \\
5 & -1116.570 & -1110.073 & -9.594 & 3.097 & 12 \\
6 & -1116.552 & -1110.166 & -9.440 & 3.054 & 12 \\
\hline
\end{tabular}
\label{table:1AJF_Dis_Malla}
\end{table}

Table \ref{table:1AJF_Dis_Malla} shows that the solvation energy is robust with respect to the distance between $\Gamma_m$ and $\Gamma_s$, with a nonlinear contribution to the energy of around -9.5 kcal/mol throughout. We then recommend using a distance $x$= 3 \AA, which is the smallest value where NC has two significant figures to -9.5. Coincidentally, this value is equivalent to the radius of one water molecule. The distance between $\Gamma_m$ and $\Gamma_s$ may depend on the charge of the specific molecule being analysed, and should be considered case-by-case.
